\documentclass[11pt, a4paper]{article}
\pdfoutput=1

\usepackage{jcappub}

\usepackage{comment}
\usepackage{amsmath}
\usepackage{amsfonts}
\usepackage{amssymb}
\usepackage{graphicx}
\usepackage{mathrsfs}
\usepackage{latexsym}
\usepackage{color}
\usepackage{slashed, cancel}
\usepackage{hyperref}
\usepackage{cleveref}
\usepackage{float}
\usepackage{tikz}
\usepackage{bbold}
\usepackage{soul} 
\usepackage{mathtools}
\usepackage{support-caption} 
\usepackage{subcaption}
\usepackage{enumitem}
\usepackage{journals}
\allowdisplaybreaks

\usepackage[utf8]{inputenc} 
\usepackage[version=3]{mhchem}
\hypersetup{urlcolor=blue, colorlinks=true} 

\usepackage{fancyhdr}
\renewcommand{\arraystretch}{1}
\numberwithin{equation}{section}

\definecolor{orange}{rgb}{1,0.4,0}
\definecolor{green}{rgb}{0,0.65,0}
\definecolor{rossos}{rgb}{0.8,0.2,0.3}
\definecolor{bluscuro}{rgb}{0.15, 0.2, .85}
\definecolor{bluchiaro}{cmyk}{1,.3,0.,0.1}
\hypersetup{colorlinks, citecolor=bluscuro, linkcolor=bluscuro, urlcolor=bluscuro}

\makeatother   
\newcommand{\GeV}{{\rm \,GeV}}

\newcommand{\MeV}{{\rm \,MeV}}
\newcommand{\keV}{{\rm \,keV}}
\newcommand{\eV}{{\rm \,eV}}

\newcommand{\cm}{{\rm \,cm}}
\newcommand{\fm}{{\rm \,fm}}
\newcommand{\km}{{\rm \,km}}
\newcommand{\s}{{\rm \,s}}
\newcommand{\g}{{\rm \,g}}
\newcommand{\K}{{\rm \,K}}

\newcommand{\Gyr}{{\rm \,Gyr}}

\newcommand{\Msun}{M_\odot}
\newcommand{\Mstar}{M_\star}
\newcommand{\Rstar}{R_\star}
\newcommand{\vstar}{v_\star}
\newcommand{\tstar}{t_\star}
\newcommand{\Tstar}{T_\star}

\newcommand{\muFn}{\mu_{F,n}}
\newcommand{\muFp}{\mu_{F,p}}
\newcommand{\muFe}{\mu_{F,e}}
\newcommand{\muFmu}{\mu_{F,\mu}}
 \newcommand{\muFi}{\mu_{F,i}}
  \newcommand{\muFl}{\mu_{F,\ell}}

\newcommand{\fFD}{f_{\rm FD}}

\newcommand{\sigmathl}{\sigma^{th}_{\ell\chi}}
\newcommand{\sigmath}{\sigma_{th}}
\newcommand{\mstar}{m_\ell^*}

\newcommand{\erf}{{\rm \,Erf}}
\newcommand{\Msq}{|\overline{M}|^2}

 \def\be   {\begin{equation}}   \def\ee   {\end{equation}}
 \def\ba   {\begin{array}}      \def\ea   {\end{array}}
 \def\bea  {\begin{eqnarray}}   \def\eea  {\end{eqnarray}}
 \def\bean {\begin{eqnarray*}}  \def\eean {\end{eqnarray*}}
 \def\nn{\nonumber}

\begin{document}


\title{Improved Treatment of Dark Matter Capture in Neutron Stars II:  Leptonic  Targets}

\author[a]{Nicole F.\ Bell,}
\author[b]{Giorgio Busoni,}
\author[a]{Sandra Robles}
\author[a]{and Michael Virgato}

\affiliation[a]{ARC Centre of Excellence for Dark Matter Particle Physics, \\
School of Physics, The University of Melbourne, Victoria 3010, Australia}
\affiliation[b]{Max-Planck-Institut fur Kernphysik, Saupfercheckweg 1, 69117 Heidelberg, Germany}

\emailAdd{n.bell@unimelb.edu.au}
\emailAdd{giorgio.busoni@mpi-hd.mpg.de}
\emailAdd{sandra.robles@unimelb.edu.au}
\emailAdd{mvirgato@student.unimelb.edu.au}

\abstract{
Neutron stars harbour matter under extreme conditions, providing a unique testing ground for fundamental interactions. We recently developed an improved treatment of dark matter (DM) capture in neutron stars that properly incorporates many of the important physical effects, and outlined useful analytic approximations that are valid when the scattering amplitude is independent of the centre of mass energy. We now extend that analysis to all interaction types. We also discuss the effect of going beyond the zero-temperature approximation, which provides a boost to the capture rate of low mass dark matter, and give approximations for the dark matter up-scattering rate and evaporation mass. We apply these results to scattering of dark matter from leptonic targets, for which a correct relativistic description is essential. We find that the potential neutron star sensitivity to DM-lepton scattering cross sections greatly exceeds electron-recoil experiments, particularly in the sub-GeV regime, with a sensitivity to sub-MeV DM well beyond the reach of future terrestrial experiments. 
}

\maketitle

\section{Introduction}


The quest to identify the cosmological dark matter (DM) is one of the forefront goals of modern science. In recent years, terrestrial dark matter direct detection experiments, which search for nuclear or electron recoil signals, have provided increasingly sensitive limits on the strength of dark matter interactions with regular matter.  These experiments are limited, however, by the size of the detector target mass that can be practically realized. For this reason, it makes sense to consider alternative targets with which dark matter can interact, such as stars and planets.  If the interaction of dark matter with these objects could be detected, they would offer a highly sensitive probe of the interaction strength, because the drawback of having to deal with uncertain astrophysical inputs is more than compensated for by the enormous target mass.

The capture of dark matter in the Sun~\citep{Gould:1987ju,Gould:1987ir,Jungman:1995df,Kumar:2012uh,Kappl:2011kz,Busoni:2013kaa,Bramante:2017xlb,Garani:2017jcj,Busoni:2017mhe} or the Earth~\citep{Gould:1987ir} has long been used as a dark matter indirect detection technique. This is achieved by searching for the annihilation of accumulated dark matter either to neutrinos~\citep{Tanaka:2011uf,Choi:2015ara,Adrian-Martinez:2016ujo,Adrian-Martinez:2016gti,Aartsen:2016zhm} or, more recently, to other dark sector particles which escape the Sun~\citep{Batell:2009zp,Schuster:2009au,Bell:2011sn,Feng:2016ijc,Leane:2017vag}.  In addition, energy transport in the Sun may be altered by the presence of DM~\citep{Gould:1989hm, Gould:1989ez, Vincent:2013lua, Geytenbeek:2016nfg}. Dark matter capture in neutron stars (NSs)~\citep{Goldman:1989nd} is particularly efficient due to the extremely high density of these objects.  Possible consequences include DM-triggered collapse of neutron stars to black holes~\citep{Goldman:1989nd,Kouvaris:2010jy,Kouvaris:2011fi,McDermott:2011jp,Guver:2012ba,Bell:2013xk, Bramante:2013nma,Bertoni:2013bsa,Garani:2018kkd,Dasgupta:2020mqg} or a modification to the rate of neutron star mergers~\citep{Bramante:2017ulk}. Recently, attention has focused on the heating of neutron stars that results from dark matter capture, thermalization and annihilation~\citep{Gonzalez,Baryakhtar:2017dbj,Raj:2017wrv,Bell:2018pkk,Camargo:2019wou,Bell:2019pyc,Garani:2019fpa,Acevedo:2019agu,Joglekar:2019vzy,Joglekar:2020liw,Dasgupta:2020dik,Ilie:2020vec,Garani:2020wge}. Similar considerations can be applied to capture in other objects, such as white dwarfs~\citep{Bramante:2017xlb,Dasgupta:2019juq}, planets~\citep{Bramante:2019fhi,Leane:2020wob} or the Moon~\citep{Garani:2019rcb}.

Despite this long history, the capture of dark matter in stars has usually employed various approximation or simplifications that neglect important physics effects.  Recent attention has thus turned to more accurate evaluations of the capture rate~\cite{Garani:2018kkd,Bell:2020jou} that correctly incorporate the important physical effects.\footnote{In the case of scattering from hadronic constituents, one must also account for the fact that the hadrons cannot be treated as point particles, and experience strong interactions in the NS medium~\cite{Bell:2020obw}. These effects are not relevant for the leptonic targets considered here.} In a recent paper, ref.~\cite{Bell:2020jou}, we provided a realistic calculation that correctly incorporated gravitational focusing, a fully relativistic scattering treatment, Pauli blocking, NS opacity and multi-scattering effects. Furthermore, in addition to providing exact expressions for the numerical evaluation of the capture rate, we also derived simplified expressions that greatly increase the computational efficiency, valid for particular interaction types.  Specifically, the approximations of ref.~\cite{Bell:2020jou} are valid when the differential cross section depends on powers of the Mandelstam variable $t$, but not on the centre of mass energy $s$. Those results were formulated for scattering on neutron targets, the most abundant NS constituent, though can be readily adapted for any target species.

In the present paper we complete and extend the results of ref.~\cite{Bell:2020jou} as outlined below:
\begin{itemize}
\item
We adapt our result to lepton targets. Importantly, note that a fully relativistic scattering treatment is essential for scattering from the highly degenerate, relativistic, electrons in the NS interior.
\item
We present analytic approximations for the DM interaction rate in NSs that are valid for $s$-dependent scattering amplitudes. Together with the result of our previous work, ref.~\cite{Bell:2020jou}, this allows efficient calculation of the DM capture rate for all possible interaction types.
\item
We calculate the capture rates for scattering from both muon and electron targets, incorporating the full radial dependence of the density and chemical potentials for these species. This allows us to determine the NS  sensitivity to DM-lepton couplings, for all interaction types. We also compare our results with other recent calculations in the literature.  
\end{itemize}

We find that the capture rate due to scattering on muons can dominate over the scattering on electrons, for some interactions types.  This is particularly so for scalar or pseudoscalar interactions, for which the DM-lepton couplings scale as the lepton mass. For other interaction types, capture due to scattering on electrons and muons is comparable, despite the lower muon abundance, due to the larger muon mass and lower muon chemical potential.

The outline of this paper is as follows:  In Section~\ref{sec:ns} we outline the relevant NS star properties and benchmark Equations of State (EoS), and determine the radial number density and chemical potential profiles for the particle species present in the star by solving the Tolman-Oppenheimer-Volkoff equations coupled to the EoS fits.  We discuss the capture rate calculation in Section~\ref{sec:capintrates}, and present useful analytic approximations for this rate and the evaporation rate in Appendices~\ref{sec:intrates} and \ref{sec:intevaprate}, respectively.  Our results are presented in Section~\ref{sec:results} and our conclusions in Section~\ref{sec:conclusions}.

\section{Neutron Stars}
\label{sec:ns}

Neutron stars (NSs), one of the possible end-points of giant stars, are  the densest stars known. They provide a unique environment to test fundamental properties of matter under extreme conditions. Even though our understanding of these objects has improved in the recent years, in the light of major theoretical and observational breakthroughs, there are still many uncertainties regarding NS composition and internal structure.
In what follows, we summarise the NS structure equations.

\subsection{Internal structure}
\label{sec:NSmodels}

NSs are primarily composed of  degenerate nuclear matter.  Several layers and phase transitions can be found in their interior from a thin atmosphere up to the innermost core layer. 
The locally homogeneous core accounts for 
$\sim 99\%$ of the NS mass \cite{Haensel:2007yy,Chamel:2008ca}.  
The outermost layer, the crust, even though it is just $\sim 1\km$ thick, is the place where thermal conduction occurs and is hence  responsible for the temperature drop between the core and the surface. 

The outer crust is made of ionised heavy elements in a Coulomb lattice and strongly degenerate electrons (similar to a white dwarf). At densities $\rho\gtrsim10^6 \g\cm^3$, the electron gas is already ultrarelativistic. 
 Moving from the surface towards the star interior, the increasing density induces electron capture and the nuclei become more and more neutron rich until 
the neutron drip density $\rho_{\rm ND}\sim 4.3 \times 10^{11} {\g \cm^{-3}}$ is reached~\cite{Ruester:2005fm,RocaMaza:2008ja,Pearson:2011,Kreim:2013rqa,Chamel:2015oqa}. This density marks the onset of the inner crust, where free neutrons, dripped off neutron rich nuclei, coexist with  neutron-proton clusters and electrons. 
At about  half the nuclear saturation density ($\rho_0=2.8 \times 10^{14} {\g \cm^{-3}}$), the nucleon clusters dissolve into their constituents as we cross the crust-core interface. 

Matter in the outer core is mainly composed of neutrons in a superfluid liquid state and an admixture of protons and electrons in beta equilibrium. Muons appear when the electron chemical potential reaches the muon mass, replacing a fraction of the electrons in order to minimise the energy of the system. This system in equilibrium is called $npe\mu$ matter and is the minimal model for the NS core. The outer core extends to densities of $\sim2\rho_0$. 
The composition of NSs at higher densities is less understood; the inner core may contain meson condensates, hyperons  or deconfined quark matter~\cite{Haensel:2007yy,Weber:2006ep,Baym:2017whm}. The appearance of these exotic species depends on the mass of the star. 
In this paper, we will focus only on neutron stars made of only $npe\mu$ matter and we will consider DM scattering off leptonic targets.

\subsection{Equation of state}

With the sole exception of the outermost crust layers (which are only a few meters thick), NS matter is mainly in a strongly degenerate state. A consequence of this is that the pressure is independent of temperature. As a result, the equation of state (EoS) of dense matter depends only on one parameter, frequently taken to be  the baryon number density, $n_b$. 
The EoS is the  key ingredient needed to solve the NS structure equations. Nevertheless, its precise determination is still one of the key open problems in nuclear astrophysics. The EoS governing the NS core is particularly challenging, even if we assume that only $npe\mu$ matter is  present, since it requires knowledge of the behaviour of strong interactions in superdense matter.

Of the several EoSs  found in the literature, see e.g. refs.~\citep{Akmal:1998cf,RikovskaStone:2006ta,Goriely:2010bm,Kojo:2014rca,Baym:2017whm,Annala:2019eax}, we  consider  the unified equations of state for cold non-accreting matter with Brussels-Montreal functionals  BSk19, BSk20, BSk21, BSk22, BSk24, BSk25 and BSk26~\cite{Goriely:2010bm,Pearson:2011,Pearson:2012hz,Goriely:2013}. 
A unified equation of state provides a thermodynamically consistent description of a NS from the surface to the core centre. 
These unified EoSs assume that a NS is made of neutrons, protons, electrons and muons, neglecting the presence of exotic matter.  Analytic fits for these EoSs are given in ref.~\cite{Potekhin:2013qqa,Pearson:2018tkr}~\footnote{These fits are also publicly available as \texttt{FORTRAN} subroutines  at \url{http://www.ioffe.ru/astro/NSG/BSk/}.}. 
These  fits not only provide us with an excellent tool for evaluating NS microscopic properties without directly performing the nuclear physics calculations, but also are easily coupled to 
the Tolman-Oppenheimer-Volkoff (TOV)  equations~\cite{Tolman:1939jz,Oppenheimer:1939ne} to obtain the stellar structure.

These EoS families were obtained under beta equilibrium. Inverse beta decay equilibrium and charge neutrality dictates the exact abundances $Y_i$ and chemical potentials $\muFi$ of the NS constituents throughout the stellar interior, 
\begin{eqnarray}
\muFn(n_b,Y_p) &=& \muFp(n_b,Y_p) +  \muFe(n_b,Y_e), \quad
\qquad \muFe(n_b,Y_e)  = \muFmu(n_b,Y_\mu), \\ 
Y_p(n_b)&=&Y_e(n_b)+Y_\mu(n_b),
\end{eqnarray}
where $Y_n(n_b)=1-Y_p(n_b)$. 
Analytic fits for these quantities in the core and the crust as a function of the baryon number density are provided in ref.~\cite{Pearson:2018tkr}.

\subsection{Benchmark models}

In addition to QCD at high density, the NS internal structure is determined by general relativity (GR). 
Therefore, to obtain radial profiles of the quantities needed in our analysis, we assume   a non-rotating, non-magnetized, spherically symmetric NS, and couple the EoS, $P=P(n_b)$, $\rho=\rho(n_b)$, to the TOV equations \cite{Tolman:1939jz,Oppenheimer:1939ne}
\begin{eqnarray}
    \frac{dP}{dr}&=&- \rho(r)c^2\left[1+\frac{P(r)}{\rho(r)c^2}\right]\frac{d\Phi}{dr}, \label{eq:tov1}\\
    \frac{d\Phi}{dr}&=&\frac{G M(r)}{ c^2r^2}\left[1+\frac{4\pi P(r) r^3 }{M(r) c^2}\right]\left[1-\frac{2GM(r)}{ c^2 r}\right]^{-1},     
 \label{eq:tov2}
\end{eqnarray}
and the mass equation
\begin{equation}
    \frac{dM}{dr}=4\pi r^2 \rho(r),
 \label{eq:NSmass}
\end{equation}
where $M(r)$ is the mass contained within a sphere of radius $r$, $\Phi(r)$ is the gravitational potential, and the Schwarzschild  metric is 
\begin{equation}
ds^2= -d\tau^2 = -B(r) c^2 dt^2 + A(r)dr^2 +r^2 d\Omega^2,     
\end{equation}
with
\begin{eqnarray}
A(r) &=& \left[1-\frac{2GM(r)}{c^2 r}\right]^{-1}, \label{eq:Arequation}\\
B(r) &=& e^{2\Phi(r)},\\
\frac{d}{dr}B(r) &=& \frac{2G}{c^2 r^2}\left[M(r)+\frac{4\pi}{c^2}P(r)r^3\right] \left[1-\frac{2GM(r)} {c^2 r}\right]^{-1} B(r). \label{eq:Brequation}
\end{eqnarray}
Note that the value of $B(r)$ at the NS surface is 
 \begin{equation}
B = B(\Rstar) = 1-\frac{2GM_\star}{c^2 \Rstar}.   
\label{eq:Bsurface}
\end{equation}

Given an EoS, the differential equation system in Eqs.~\ref{eq:tov1}, \ref{eq:tov2} and \ref{eq:NSmass} 
can be solved from the NS centre where $\rho(0)=\rho_c$, with $\rho_c$ a free parameter, to the  outermost layer of the crust where $\rho=10^6\g\cm^{-3}$. At that density the NS radius, $\Rstar$, and the gravitational mass of the star $\Mstar=M(r=\Rstar)$ are determined.

\begin{figure}
    \centering
    \includegraphics[width=0.6\textwidth]{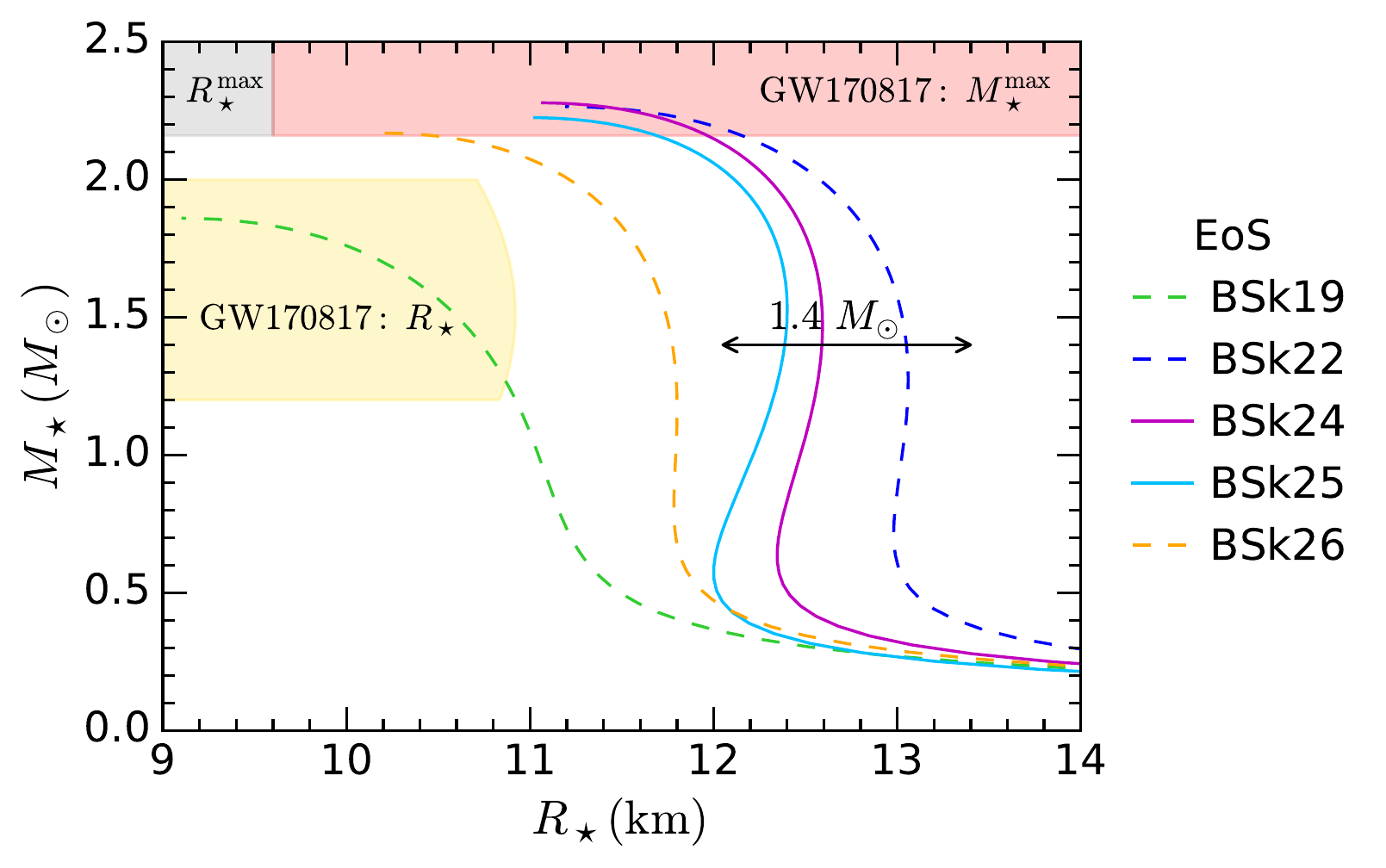}
    \caption{Mass radius relation for the functionals BSk19, BSk22, BSK24, BSk25 and BSk26. Shaded regions denote constraints on the NS maximum mass (red) \cite{Margalit:2017dij,Shibata:2017xdx,Ruiz:2017due,Rezzolla:2017aly,Shibata:2019ctb} and radius (grey) \cite{Bauswein:2017vtn} from the NS binary merger GW170817. The shaded yellow region represents the lower bound on the NS radius  derived in ref.~\cite{Koppel:2019pys} from the analysis of the GW170817 event. The $2\sigma$ confidence level constraint on the radius of a $1.4\, \Msun$ NS is shown in black \cite{Most:2018hfd}. Dashed lines denote EoS families excluded by observations. }
    \label{fig:BSKfunctionals}
\end{figure}

In Fig.~\ref{fig:BSKfunctionals}, we show the mass radius relation for the above mentioned BSk functionals. 
We do not consider BSk20 and BSk21 since they yield very similar results to BSk26 and BSk24, respectively~\cite{Perot:2019gwl}. In addition, the functionals BSk19-21 were fitted to older atomic mass data than the new series of functionals BSk22, BSk24, BSk25 and BSk26. BSk19 fails to accommodate massive NSs  and part of its parameter space is ruled out by the lower bound on the NS radius inferred from  observations of the electromagnetic counterpart of the NS binary merger event GW170817 \cite{Koppel:2019pys}. BSk22 is ruled out by constraints on the tidal deformability parameter also from GW170817 \cite{Perot:2019gwl}, and because it requires direct Urca processes operating in most NSs~\cite{Pearson:2018tkr}. On the other hand, direct Urca processes are not allowed in  stable NSs described by BSk26, which contradicts observations \cite{Pearson:2018tkr}. BSk24 and BSk25 agree with current neutron star cooling observations~\cite{Chamel:2019hml}, with BSk24 giving slightly better NS mass fits to  observational data \cite{Pearson:2018tkr}. 
Therefore, as in refs.~\cite{Bell:2019pyc,Bell:2020jou}, we select the BSk24 functional.  

\begin{figure}[t] 
\centering
\includegraphics[width=0.495\textwidth]{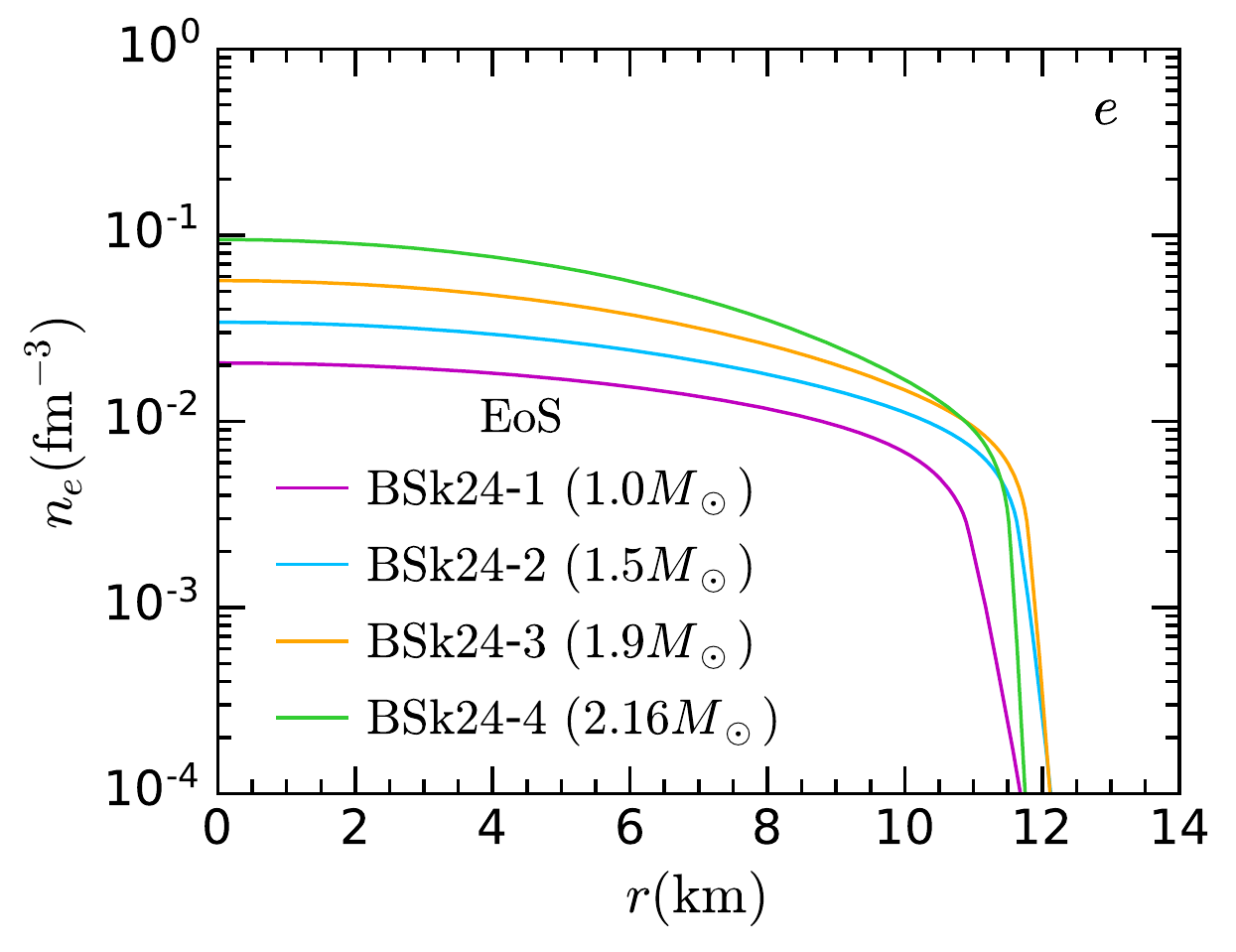}
\includegraphics[width=0.495\textwidth]{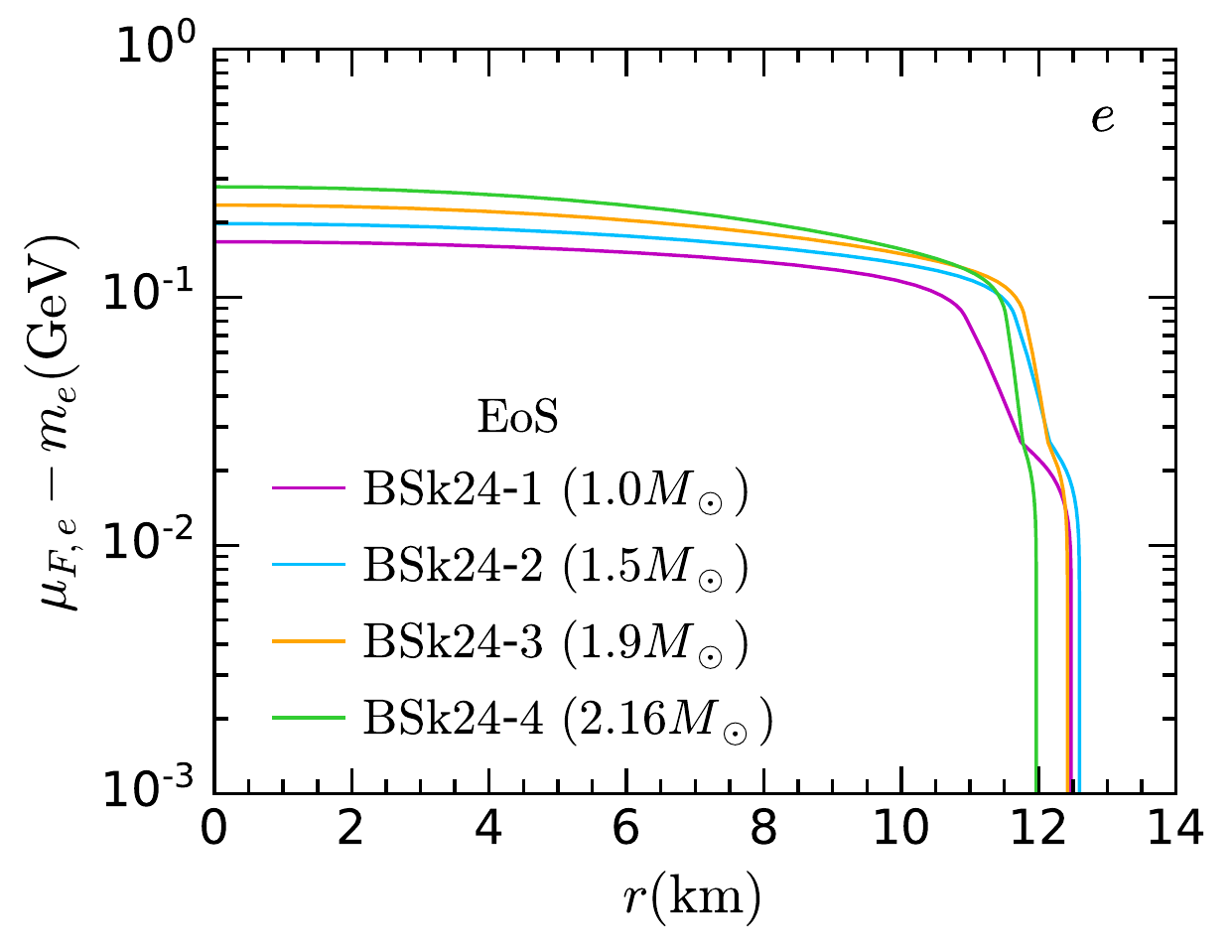}
\includegraphics[width=0.495\textwidth]{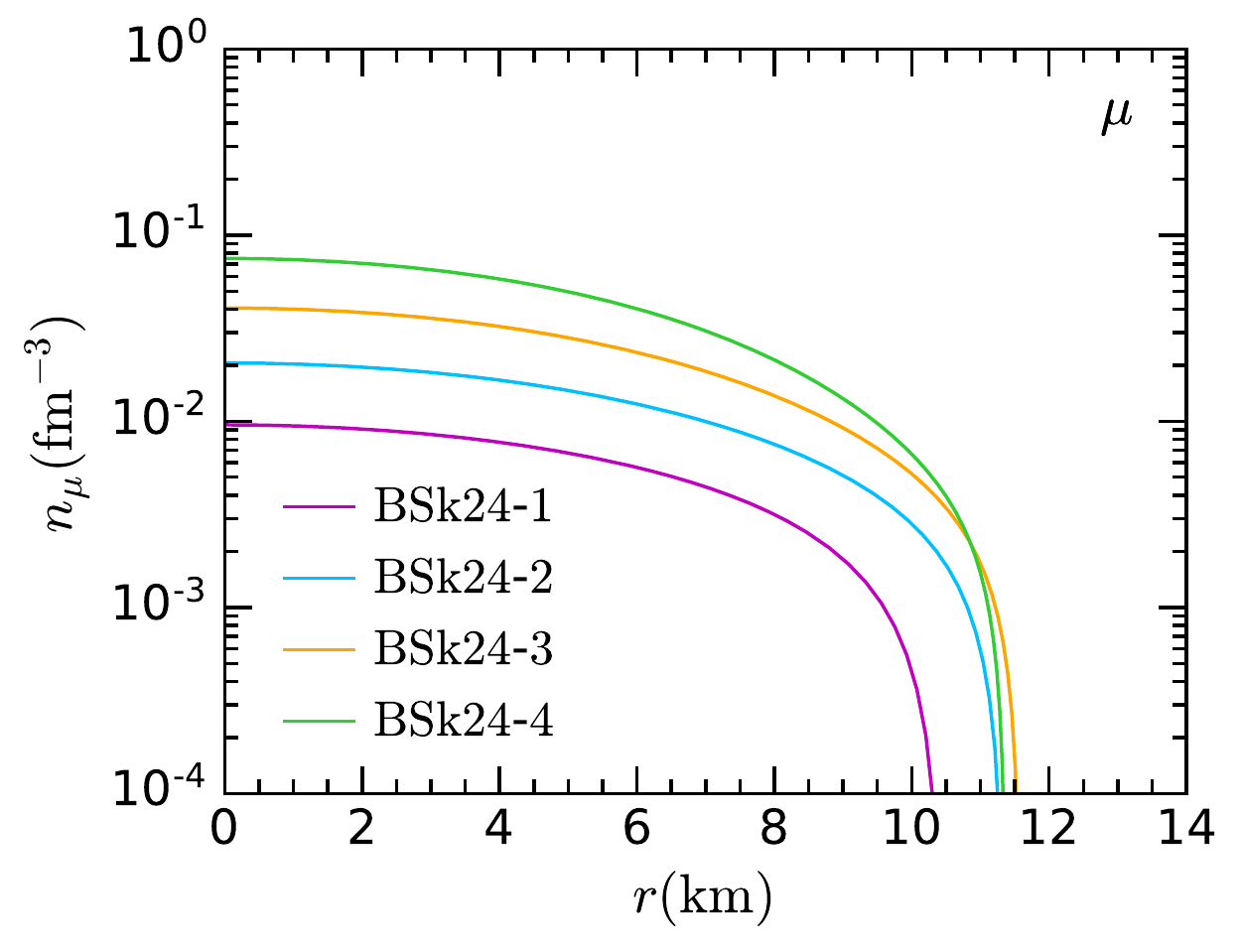}
\includegraphics[width=0.495\textwidth]{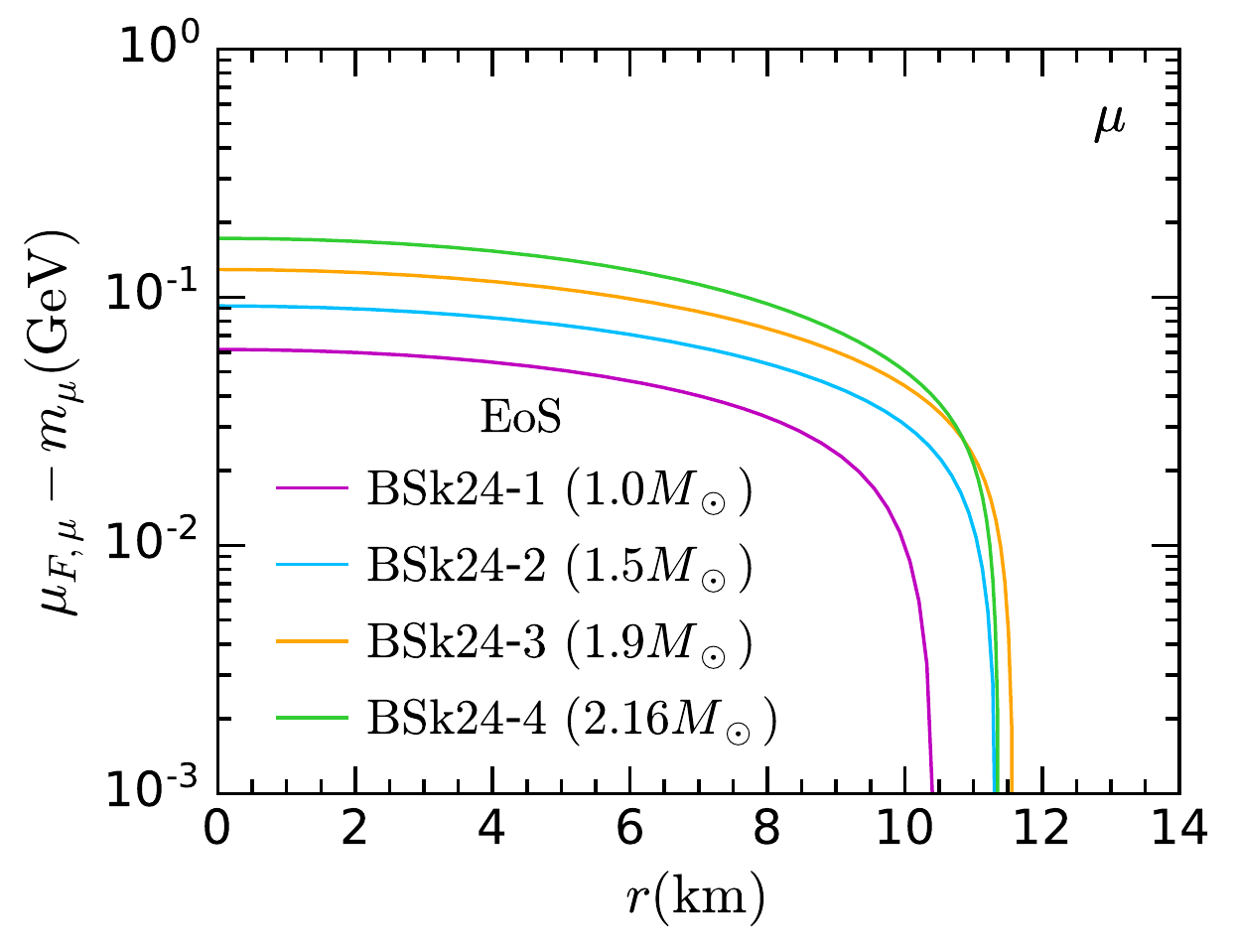}
\caption{Number density profile (left) and chemical potential (right) for electrons (top) and muons (bottom) and NS configurations of the  BSk24 functional in Table~\ref{tab:eos}. 
}
\label{fig:NSradprofs1}
\end{figure}

Coupling the BSk24 functional precision fits to the TOV equations~\ref{eq:tov1}-\ref{eq:NSmass}, we solve the differential equation system from the core centre out to the outer crust, at every radial step we determine   chemical potentials and particle number fractions for the different species using the appropriate functions for the core and the crust, available in ref.~\cite{Pearson:2018tkr}. 
We also calculate GR corrections encoded in the $B$(r) profile.
We have thus obtained radial profiles for   particle number densities $n_i$ and chemical potentials $\muFi$ for every NS constituent.  These profiles  depend on the EoS choice, i.e. on the initial parameter  $\rho_c$. We have chosen the same four configurations of the functional BSk24  given in ref.~\cite{Bell:2020jou}, see Table~\ref{tab:eos}, where the NS mass range is motivated by observations \cite{Ozel:2016oaf,Antoniadis:2016hxz} and the maximum NS mass considered is limited  by the GW170817 event to $\Mstar\lesssim2.16\Msun$ \cite{Margalit:2017dij,Shibata:2017xdx,Ruiz:2017due,Rezzolla:2017aly,Shibata:2019ctb}\footnote{More conservative limits on the maximum NS mass can be imposed by combining all the studies above~\cite{Perot:2019gwl}.}. 
In Fig.~\ref{fig:NSradprofs1} we plot the corresponding lepton profiles.
As mentioned above, electrons are present in the core and the crust while muons appear at baryon number densities $n_b\simeq 0.12\fm^{-3}$. The kink observed in the electron chemical potential marks out the transition from the core to the inner crust. 
The aforementioned radial profiles will be used in the following section to calculate the capture rate.

  \begin{table}[tb]
\centering
\begin{tabular}{|l|c|c|c|c|}
\hline
\bf EoS & \bf BSk24-1 & \bf BSk24-2 & \bf BSk24-3 & \bf BSk24-4 \\ \hline
$\rho_c$ $[\rm{g \, cm^{-3}}]$ & $5.94 \times 10^{14}$   & $7.76 \times 10^{14}$ & $1.04 \times 10^{15}$ & $1.42 \times 10^{15}$  \\
$n_b^c$ $[\fm^{-3}]$ & 0.330 & 0.430 & 0.549 & 0.670 \\
$\Mstar$ $[\Msun]$ & 1.000 & 1.500 & 1.900 & 2.160  \\
$\Rstar$ [km] & 12.215  & 12.593 & 12.419 & 11.965  \\
$B(\Rstar)$ & 0.763 & 0.648 & 0.548 & 0.467 \\
\hline
\end{tabular} 
\caption{Benchmark NSs, made of $npe\mu$ matter, for four different  configurations of the equations of state (EoS) for cold non-accreting neutron stars with Brussels–Montreal functionals BSk24 \cite{Pearson:2018tkr}. EoS configurations are determined by the central mass-energy density $\rho_c$.}
\label{tab:eos}
\end{table}

\section{Capture and Interaction Rates}
\label{sec:capintrates}

In ref.~\cite{Bell:2020jou}, we derived general expressions for the capture and interaction rates of DM in NSs, valid for a broad range of DM masses, for arbitrary NS targets and DM-target cross sections. These expressions properly take into account relativistic kinematics, including the motion of the target particles in the NS; gravitational focusing; Pauli blocking (relevant at low DM masses); the effect of the star opacity and multiple scattering (important in the capture of heavy DM); and the NS internal structure. In that which follows, we summarise and extend those results.  

\subsection{Interaction Rate}
\label{sec:intratetext}

The DM scattering rate as a function of arbitrary DM energy, and the corresponding differential interaction rate, are the key elements of the capture calculations. 
In addition, they are a necessary input in constructing the probability density function of the DM energy loss.  This is required to define the capture probability after $N$ scatterings~\cite{Bell:2020jou}, which is relevant to the capture of heavy DM via multiple scattering. 

Following refs.~\citep{Bertoni:2013bsa} and \cite{Bell:2020jou}, we define the  DM scattering rate as
\begin{eqnarray}
\Gamma &=& \int \frac{d^3k^{'}}{(2\pi)^3} \frac{1}{(2E_\chi)(2E^{'}_\chi)(2m_\ell)(2m_\ell)}\Theta(E^{'}_\chi-m_\chi)\Theta(q_0)S(q_0,q), \label{eq:intratedeftext}\\
S(q_0,q) &=& 2\int \frac{d^3p}{(2\pi)^3}\int \frac{d^3p^{'}}{(2\pi)^3} \frac{m_i^2}{E_\ell E^{'}_\ell}|\overline{M}|^2 (2\pi)^4\delta^4\left(k_\mu+p_\mu-k_\mu^{'}-p_\mu^{'}\right)\nonumber\\
 &&\times\fFD(E_\ell)(1-\fFD(E^{'}_\ell))\Theta(E_\ell-m_\ell)\Theta(E^{'}_\ell-m_\ell),
 \label{eq:responsefunc}
\end{eqnarray}
where $k^\mu=(E_\chi$,$\vec{k})$, $k^{'\mu}=(E^{'}_\chi,\vec{k'})$ are the DM initial and final momenta, $p^\mu=(E_\ell,\vec{p})$ and $p^{'\mu}=(E^{'}_\ell,\vec{p'})$ are the target particle initial and final momenta, $m_\ell$ is the lepton target mass, $q_0=E_\ell^{'}-E_\ell$ is the DM energy loss and $\fFD$ is the Fermi Dirac distribution. We define the quantity $S(q_0,q)$ to be the response function, which contains the dependence on the squared matrix element, $\Msq$. To calculate the interaction rate,  
we consider the interaction of Dirac DM with SM leptons, described by the dimension 6 effective operators listed in Table~\ref{tab:operatorshe}, where the strength of the coupling is parametrised by the cutoff scale $\Lambda$ and $\mu=m_\chi/m_\ell$. 

In ref.~\cite{Bell:2020jou}, we showed that Eq.~\ref{eq:intratedeftext} can be solved analytically for DM-nucleon differential cross sections that depend on powers of the Mandelstam variable $t$ (but not on $s$) as a function of the DM energy $E_\chi$. 
In that case, assuming that $\Tstar\rightarrow0$, Eq.~\ref{eq:intratedeftext} reduces to
\begin{eqnarray}
\Gamma^{-}(E_\chi) &\propto& \frac{1}{2^7\pi^3E_\chi k }\int_0^{E_\chi-m_\chi}q_0 dq_0 \int \frac{t_E^n dt_E }{\sqrt{q_0^2+t_E}}  \left[1-g_0\left(\frac{E_\ell^{\,t^{-}}-\muFl}{q_0}\right)\right],
\label{eq:gammafinaltext}
\end{eqnarray}
 where $t_E=-t=q^2-q_0^2$ and
 $\muFl$ is the target chemical potential after subtracting the target rest mass energy as in Fig.~\ref{fig:NSradprofs1}. From here on and as in ref.~\cite{Bell:2020jou}, we denote by $\muFl$ the Fermi energy (without the rest mass) of the leptonic species $\ell$. The quantity
\begin{equation}
E_\ell^{\, t^{-}} = -\left(m_\ell+\frac{q_0}{2}\right) + \sqrt{\left(m_\ell+\frac{q_0}{2}\right)^2+\left(\frac{\sqrt{q^2-q_0^2}}{2}-\frac{m_\ell q_0}{\sqrt{q^2-q_0^2}}\right)^2}, 
\end{equation}
is the minimum energy of the target before the collision, which is obtained from kinematics,  and $g_0(x)$ is a step function with a smooth transition, 
\begin{align}
g_0(x) =  \begin{cases}
\, 1 \quad &x>0, \\
\, 1+x \quad &-1<x<0,\\
\, 0 \quad &x<-1.
\end{cases}
\end{align} 
The explicit integrals over $t_E$ are given in appendix~B of ref.~\cite{Bell:2020jou}. 
The differential interaction rate $\frac{d\Gamma}{d q_0}(E_\chi,q_0)$ is the integrand of Eq.~\ref{eq:gammafinaltext}. 
These expressions  can be applied to squared matrix elements that depend on linear combinations of $t^n$, with $n=0,1,2$. As seen in Table~\ref{tab:operatorshe}, this is applicable to the D1-D4 operators.

\begin{table}
\centering
{\renewcommand{\arraystretch}{1.3}
\begin{tabular}{ | c | c | c | c |}
  \hline                        
  Name & Operator & Coupling & $|\overline{M}|^2(s,t)$   \\   \hline
  D1 & $\bar\chi  \chi\;\bar \ell \ell  $ & ${y_\ell}/{\Lambda^2}$ & $\frac{y_\ell^2}{\Lambda^4} \frac{\left(4 m_{\chi }^2-t\right) \left(4 m_{\chi }^2-\mu ^2
   t\right)}{\mu ^2}$ \\  \hline
  D2 & $\bar\chi \gamma^5 \chi\;\bar \ell \ell $ & $i{y_\ell}/{\Lambda^2}$ & $\frac{y_\ell^2}{\Lambda^4} \frac{t \left(\mu ^2 t-4 m_{\chi }^2\right)}{\mu ^2}$ \\  \hline
  D3 & $\bar\chi \chi\;\bar \ell \gamma^5  \ell $&  $i{y_\ell}/{\Lambda^2}$ &  $\frac{y_\ell^2}{\Lambda^4} t \left(t-4 m_{\chi }^2\right)$ \\  \hline
  D4 & $\bar\chi \gamma^5 \chi\; \bar \ell \gamma^5 \ell $ & ${y_\ell}/{\Lambda^2}$  & $\frac{y_\ell^2}{\Lambda^4} t^2$ \\  \hline
  D5 & $\bar \chi \gamma_\mu \chi\; \bar \ell \gamma^\mu \ell$ & ${1}/{\Lambda^2}$ &  $2\frac{1}{\Lambda^4} \frac{2 \left(\mu ^2+1\right)^2 m_{\chi }^4-4 \left(\mu ^2+1\right) \mu ^2 s m_{\chi }^2+\mu ^4 \left(2 s^2+2 s t+t^2\right)}{\mu^4}$ \\  \hline
  D6 & $\bar\chi \gamma_\mu \gamma^5 \chi\; \bar  \ell \gamma^\mu \ell $ & ${1}/{\Lambda^2}$ &  $2\frac{1}{\Lambda^4} \frac{2 \left(\mu ^2-1\right)^2 m_{\chi }^4-4 \mu ^2 m_{\chi }^2 \left(\mu ^2 s+s+\mu ^2 t\right)+\mu ^4 \left(2 s^2+2 s
   t+t^2\right)}{\mu^4}$  \\  \hline
  D7 & $\bar \chi \gamma_\mu  \chi\; \bar \ell \gamma^\mu\gamma^5  \ell$ & ${1}/{\Lambda^2}$ &  $2\frac{1}{\Lambda^4} \frac{2 \left(\mu ^2-1\right)^2 m_{\chi }^4-4 \mu ^2 m_{\chi }^2 \left(\mu ^2 s+s+t\right)+\mu ^4 \left(2 s^2+2 s t+t^2\right)}{\mu^4}$ \\  \hline
  D8 & $\bar \chi \gamma_\mu \gamma^5 \chi\; \bar \ell \gamma^\mu \gamma^5 \ell $ & ${1}/{\Lambda^2}$ &  $2\frac{1}{\Lambda^4} \frac{2 \left(\mu ^4+10 \mu ^2+1\right) m_{\chi }^4-4 \left(\mu ^2+1\right) \mu ^2
   m_{\chi }^2 (s+t)+\mu ^4 \left(2 s^2+2 s t+t^2\right)}{\mu ^4}$ \\  \hline
  D9 & $\bar \chi \sigma_{\mu\nu} \chi\; \bar \ell \sigma^{\mu\nu} \ell $ & ${1}/{\Lambda^2}$ & $8\frac{1}{\Lambda^4} \frac{4 \left(\mu ^4+4 \mu ^2+1\right) m_{\chi }^4-2 \left(\mu ^2+1\right) \mu ^2 m_{\chi
   }^2 (4 s+t)+\mu ^4 (2 s+t)^2}{\mu ^4}$  \\  \hline
 D10 & $\bar \chi \sigma_{\mu\nu} \gamma^5\chi\; \bar \ell \sigma^{\mu\nu} \ell \;$ & ${i}/{\Lambda^2}$ &  $8\frac{1}{\Lambda^4} \frac{4 \left(\mu ^2-1\right)^2 m_{\chi }^4-2 \left(\mu ^2+1\right) \mu ^2 m_{\chi }^2 (4 s+t)+\mu ^4 (2 s+t)^2}{\mu^4}$ \\  \hline
\end{tabular}}
\caption{EFT operators~\cite{Goodman:2010ku} and squared matrix elements for the scattering of Dirac DM from leptons. The coefficient of each operator is given as a function of the lepton Yukawa coupling, $y_\ell$, and the cutoff scale, $\Lambda$. The fourth column shows the squared matrix elements at high energy as a function of the Mandelstam variables $s$ and $t$. 
\label{tab:operatorshe} }
\end{table}

For the remaining operators D5-D10, we require either a numerical computation or an analytical approach that generalises that of ref.~\cite{Bell:2020jou} to now handle $s$-dependent interaction rates. We derive such analytical expressions for $s$-dependent interaction rates for the first time, with our results presented in appendix~\ref{sec:intrates}. It is worth noting that these expressions are  valid only in the zero temperature approximation. 
With these results, the interaction rates for operators D5-D10 can be obtained as linear combinations of those for simple power laws $|\overline{M}|^2\propto t^n s^m$. There are 6 possible power laws in total, namely  $1,t,t^2,s,st,s^2$. 
The methodology to calculate the full expressions for $\Msq\propto t^n s^m$ is similar to that adopted for $s$-independent matrix elements in ref.~\cite{Bell:2020jou}, with a few additions that are outlined in appendix~\ref{sec:intrates}. We do not report the full expressions for $\Msq\propto t^n s^m$ due to their length.

\subsection{Capture Rate}
\label{sec:capture}

Below we provide a summary of the various expressions for the capture rate and the regimes for which they apply, the details of which can be found in ref.~\cite{Bell:2020jou}.

\begin{enumerate}
\item
{\bf Optically thin, single scatter:  $\sigma\ll\sigmathl$ and $m_\chi\lesssim \mstar$} \\
We begin by defining $c_1$ to be the probability that a single scattering interaction will result in capture of the DM particle.
The simplest regime occurs when the cross section is much smaller than the threshold cross section, $\sigma\ll\sigmathl$, and the DM mass is much smaller than $\mstar$, where $\mstar$ is the DM mass for which multiple scattering becomes relevant. Both quantities $\sigmathl$ and $\mstar$ depend on the specific target $i$. To calculate $\mstar$ for operators D1-D10, we use the differential interaction rates $\frac{d\Gamma}{d q_0}$ computed in section~\ref{sec:intratetext} and follow the approach outlined in  ref.~\citep{Bell:2020jou}.
In Table~\ref{tab:mstarsigma}, we show typical values of $\mstar$ and $\sigmathl$ for electron and muon targets.  Note that the exact value of $\mstar$ depends on $B(r)$,  $\muFl(r)$ and the type of interaction.

When the above mentioned conditions are met, the capture probability is of order one, $c_1\sim1$, and the neutron star can be treated as optically thin.  In this limit, the capture rate is given by
\begin{eqnarray}
C &=& \frac{4\pi}{\vstar} \frac{\rho_\chi}{m_\chi} {\rm Erf }\left(\sqrt{\frac{3}{2}}\frac{\vstar}{v_d}\right)\int_0^{\Rstar}  r^2 \frac{\sqrt{1-B(r)}}{B(r)} \Omega^{-}(r)  \, dr, \label{eq:capturefinalM2text} \\
\Omega^{-}(r) &=& \frac{\zeta(r)}{32\pi^3}\int dt dE_\ell ds |\overline{M}|^2 \frac{E_\ell}{2s\beta(s)-\gamma^2(s)} \frac{1}{p_\chi}\frac{s}{\gamma(s)}\fFD(E_\ell,r)(1-\fFD(E_\ell^{'},r)),\label{eq:intrate}
\end{eqnarray}
where $\rho_\chi$ is the local DM density, $\vstar$ is the NS velocity, $v_d$ is the DM velocity dispersion, $\zeta(r)=\frac{n_{\ell}(r)}{n_{free}(r)}$ is a correction factor due to the use of a realistic target number density $n_\ell(r)$, 
$E_\ell$ and $E_\ell^{'}$ are the initial and final energy respectively of the target $\ell$, and 
\begin{eqnarray}
    \beta(s) &=& s-\left(m_\ell^2+m_\chi^2\right),\\
    \gamma(s) &=& \sqrt{\beta^2(s)-4m_\ell^2m_\chi^2},\\
    p_\chi &=& m_\chi \sqrt{\frac{1-B(r)}{B(r)}}.\label{eq:pchi}    
\end{eqnarray}
The integration intervals for $s$, $t$ and $E_i$ are given in ref.~\cite{Bell:2020jou}. 
Note that Eq.~\ref{eq:intrate} correctly accounts for Pauli blocking, given by the $1-\fFD$ term, which, for muons, is relevant for $m_\chi\lesssim m_\mu$. Electrons, on the other hand, are ultra-relativistic throughout the inner crust and the core, with Pauli suppression effective for $m_\chi\lesssim \muFe$.

\begin{table}[tb]
\centering
\begin{tabular}{|l|c|c|}
\hline
\bf Target & \bf $\mu$ & \bf $e$  \\ \hline
$\mstar (\GeV)$  & $[0.3,3] \times 10^{5}$ & $[0.05,1.7] \times 10^{5}$  \\
$\sigmathl (\cm^2)$ &  $8\times 10^{-44}$ & $3\times 10^{-44}$ \\
\hline
\end{tabular} 
\caption{Typical values of $\mstar$ and $\sigmathl$ for lepton targets. 
The exact value of $\sigmathl$ depends on the DM mass, and the operator. We show  here the simplest case of constant matrix element; other operators give similar results. The threshold cross section is approximately constant in the range $1\GeV\lesssim m_\chi\lesssim \mstar$, and takes larger values outside that range with a $1/m_\chi$ or $m_\chi$ scaling for small and large masses, respectively. }
\label{tab:mstarsigma}
\end{table}

\item
{\bf Optically thin, large mass - multiple scattering: $\sigma\ll\sigmathl$ and $m_\chi\gtrsim \mstar$, } \\
For $m_\chi\gtrsim \mstar$ and $\sigma\ll\sigmathl$, the assumption that the capture probability is $c_1\sim1$ no longer holds. In fact, it is significantly smaller than 1.
We calculate the capture rate using the following approximation: 
\begin{eqnarray}
C_{approx}^* =\frac{4\pi}{\vstar}\frac{\rho_\chi}{m_\chi}{\rm Erf}\left(\sqrt{\frac{3}{2}}\frac{\vstar}{v_d}\right)  \int  r^2 dr  \frac{\sqrt{1-B(r)}}{B(r)}\Omega^{-}(r) \frac{1}{n^*_\ell(r)}
\label{eq:capturesimplelargem},
\end{eqnarray} 
where the capture probability is given by 
\begin{equation}
   c_1= \frac{1}{n^*_\ell} = 1-e^{-\mstar/m_\chi}\rightarrow \frac{\mstar}{m_\chi} \quad m_\chi\gg \mstar, 
\end{equation}
where $n^*_\ell$ represents the average number of interactions with target species $\ell$ required to remove a DM particle from the incoming flux. In this way, a suitable approximation that accounts for multiple scattering is obtained.  Typical values are reported in Table~\ref{tab:mstarsigma} for lepton targets.

\item
{\bf Optical depth: $\sigma\sim\sigmathl$.} \\
If $\sigma\sim\sigmathl$, the optically thin limit is not valid and hence we must modify the capture rate expressions above (Eqs.~\ref{eq:capturefinalM2text} and \ref{eq:capturesimplelargem}) to include an optical factor $\eta(r)$ \citep{Bell:2020jou}. This is an extinction factor that accounts for the star opacity.  We then have 
\begin{equation}
C_{opt} = \frac{4\pi}{\vstar}\frac{\rho_\chi}{m_\chi}{\rm Erf}\left(\sqrt{\frac{3}{2}}\frac{\vstar}{v_d}\right) \int_0^{\Rstar}  r^2 dr  \frac{\sqrt{1-B(r)}}{B(r)}\Omega^{-}(r) \eta(r).  \label{eq:captureopticaldepth}
\end{equation}

\item
{\bf Geometric limit: $\sigma\gg\sigmathl$.} \\
In this case we can safely estimate the capture rate using the geometric limit calculated in ref.~\cite{Bell:2018pkk}, 
\begin{equation}
C_{geom} =  \frac{\pi R_\star^2[(1-B(R_\star)]}{v_\star B(R_\star)} \frac{\rho_\chi}{m_\chi} \erf\left(\sqrt{\frac{3}{2}}\frac{v_\star}{v_d}\right).
\label{eq:capturegeom}    
\end{equation}
\end{enumerate}

\section{Results}
\label{sec:results}

\subsection{Capture Rate}
\label{sec:caprateresults}
 \begin{figure}[t] 
\centering
\includegraphics[width=\textwidth]{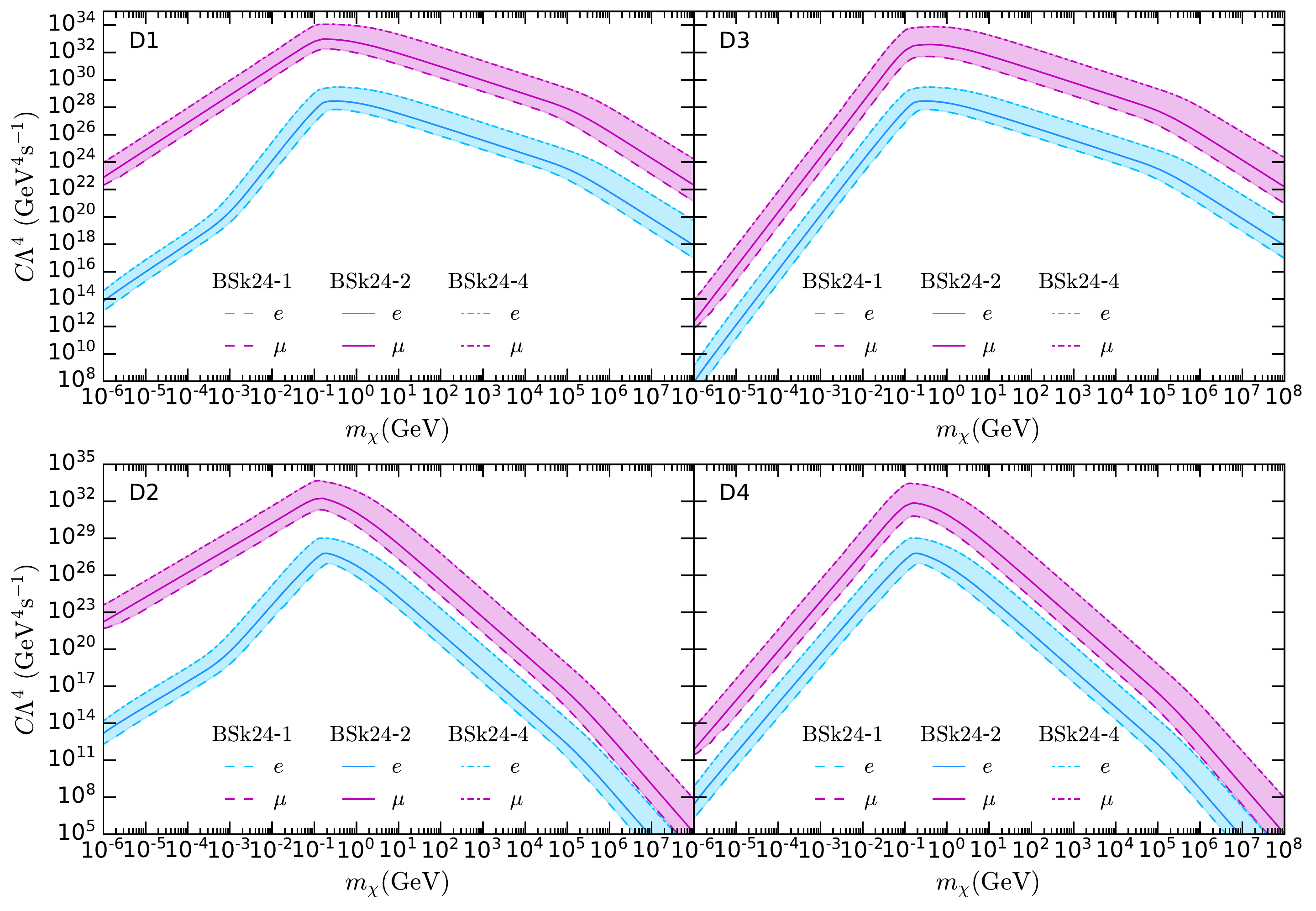}
\caption{Capture rate in the optically thin limit for operators D1-D4 as a function of the DM mass $m_\chi$ for electrons (light blue) and muons (magenta) in the NS benchmark models BSk24-1 (dashed), BSk24-2 (solid) and BSk24-4 (dot-dashed). The shaded regions denote the change in the capture rate with the NS configuration for the same EoS family BSk24. 
All capture rates scale as $\Lambda^{-4}$. We require $\Lambda$ to be sufficiently large that the capture rates are smaller than the geometric limit, $C_{geom}$. 
}
\label{fig:capratesD1D4}
\end{figure} 

 \begin{figure}[t] 
\centering
\includegraphics[width=\textwidth]{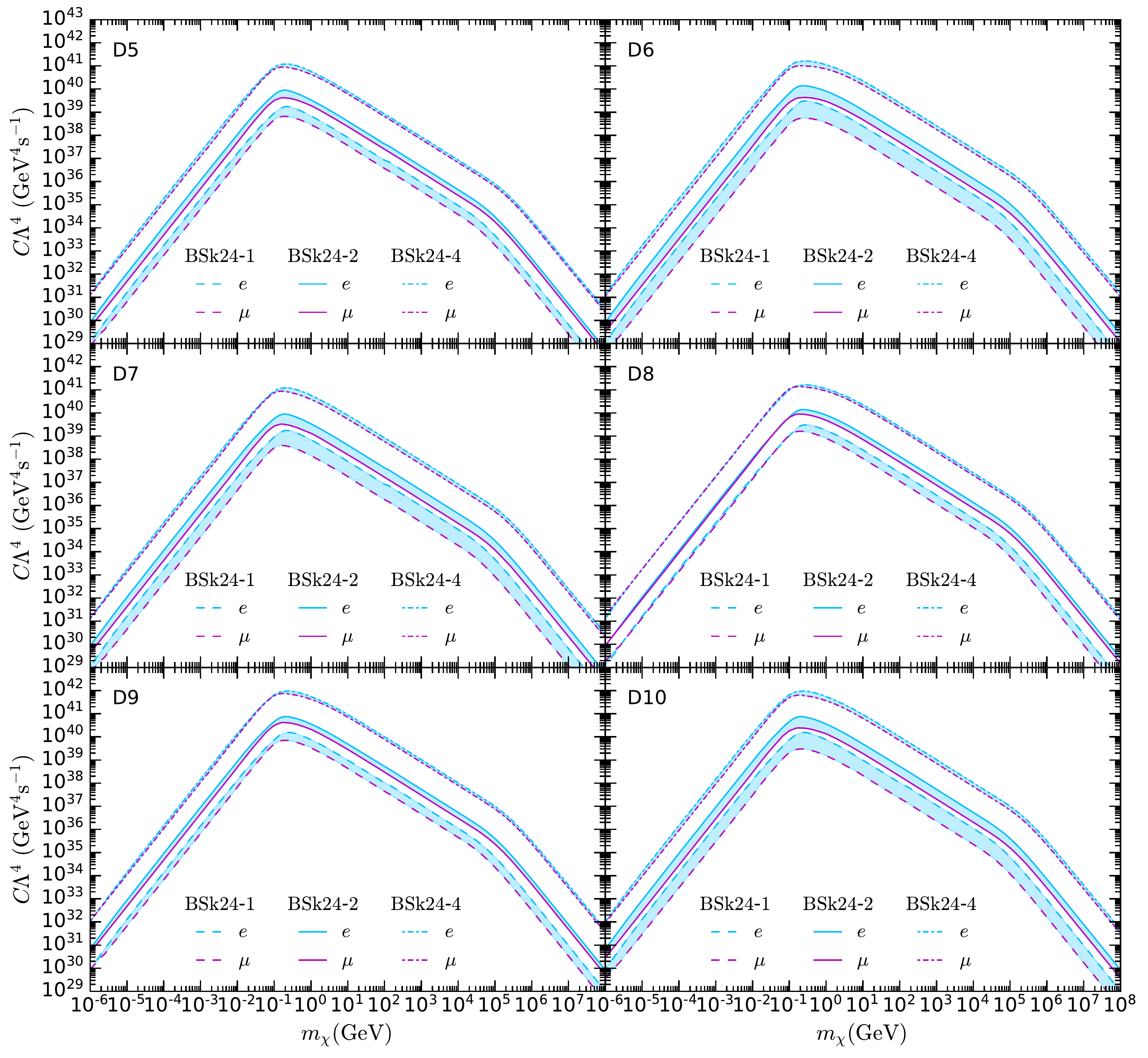}
\caption{Capture rate  in the optically thin limit for operators D5-D10 as a function of the DM mass $m_\chi$ for electrons (light blue) and muons (magenta) in the NS benchmark models BSk24-1 (dashed), BSk24-2 (solid) and BSk24-4 (dot-dashed). All capture rates scale as $\Lambda^{-4}$. The shaded regions depict the difference between capture by electrons and muons for the above mentioned NS models. 
}
\label{fig:capratesD5D10}
\end{figure}

In this section, we present results for the capture rate $C\Lambda^4$ for each of the EFT operators in Table~\ref{tab:operatorshe}, calculated in the optically thin limit using Eq.~\ref{eq:capturefinalM2text} for $m_\chi\lesssim \mstar$ and  Eq.~\ref{eq:capturesimplelargem} for $m_\chi\gtrsim \mstar$\footnote{To numerically solve these equations we use the \texttt{CUBA} libraries \citep{Hahn:2004fe,Hahn:2014fua} linked to \texttt{Mathematica}~\cite{Mathematica}.}.
Figs.~\ref{fig:capratesD1D4} and \ref{fig:capratesD5D10} show the results for electron (light blue) and muon (magenta) targets, considering three NS benchmark models: BSk24-1 (dashed, $1\Msun$), BSk24-2 (solid, $1.5\Msun$) and BSk24-4 (dot-dashed, $2.16\Msun$). In addition, we assume a nearby NS, located in the Solar neighbourhood, and thus take $\rho_\chi=0.4\GeV\cm^{-3}$, $\vstar=230\km\s^{-1}$ and $v_d=270\km\s^{-1}$.  

In these figures, we observe that the capture rate is suppressed due to Pauli blocking when $m_\chi \lesssim m_\mu$. The change of slope at $m_\chi\sim \mstar \sim 10^5\GeV$, observed for both targets, is due to multiple scattering. 
Note that the slope of the capture rate for the three distinctive regions, low mass, intermediate ($m_\mu\lesssim m_\chi\lesssim \mstar$) and large mass (multiple scattering) is very similar for operators D5-D10 (Fig.~\ref{fig:capratesD5D10}), while for D1-D4 (Fig.~\ref{fig:capratesD1D4}), the shape of $C$ is controlled by the power of $t$ that dominates the interaction, which in general is the lowest power \cite{Bell:2020jou}. 
The sole exception to this are the capture rates for operators D1 and D2 with electron targets, which show a distinctive feature in the region $m_e\lesssim m_\chi\lesssim 100\MeV$ that does not occur for the other operators.  The $C$ rate for D1 and D2 is more suppressed in that particular region, similarly to D3 and D4, respectively. This is due to the form of the  corresponding matrix elements together with the smallness of the electron mass. Namely, 
D1 and D2 are the only two operators that contain a factor $(t-4 m_\ell^2)$ in their scattering amplitudes, for electrons this means that the lowest power of $t$ in $\Msq$ is multiplied by $m_e^2$, i.e. these terms are suppressed in the $m_e \lesssim m_\chi\lesssim 100\MeV$ interval. Consequently, the capture rate in that DM mass region is dominated by the unsuppressed $t$-terms  in $\Msq$, $t$  for D1 (as for D3) and $t^2$ for D2 (see Table~\ref{tab:operatorshe}), while below $m_e$ this additional suppression disappears and the capture rate follows the lowest power of $t$ as for muon targets. 

 From Fig.~\ref{fig:capratesD1D4}, we note that for the same cutoff scale $\Lambda$, the muon contribution to the total  capture rate for operators D1-D4 surpasses that of the electron by approximately 4 orders of magnitude for most of the DM mass range, and by about 8 orders of magnitude at very low mass for operators D1-D2 (because of the additional suppression described above). This is due to the large hierarchy between DM couplings to electrons and muons, which is of order $(\frac{m_\mu}{m_e})^2$. 
Conversely, for operators D5-D10, electrons and muons have the same couplings (see Table~\ref{tab:operatorshe}). However, despite similar couplings and a lower abundance, muons are able to  capture DM at a rate comparable to electrons (see light blue regions in Fig.~\ref{fig:capratesD5D10}), thanks to their larger mass and lower chemical potential (see Fig.~\ref{fig:NSradprofs1}, right panels), i.e., their interactions with DM are less Pauli suppressed. The small difference between the rates at which electron and muon are able to capture DM particles reduces for heavier NS configurations, e.g. from a factor $\sim5$ (BSk24-1) to $\sim 1.5$ (BSk24-4) for D6 and D10; see the light blue shaded regions in Fig.~\ref{fig:capratesD5D10}. Recall that muons are expected to be found in larger fractions in massive NSs (see Fig.~\ref{fig:NSradprofs1}, left panels).

It is also worth noting that different EoS assumptions can lead to variations in the capture rate for electron targets of at least two orders of magnitude in the Pauli suppressed region and $\sim 2.5$ orders of magnitude in the large DM mass regime (compare dashed with dot-dashed light blue lines). For muons, the effect is even larger, with capture rate variations from $\sim {\cal O}(5\times 10^2)$ for low DM mass to $\sim {\cal O}(2\times 10^3)$ for heavy DM, when comparing the lightest and most massive NS configurations of the BSk24 family. For the operators D2 and D4, these variations are even more pronounced for both electrons and muons and can reach  $\sim {\cal O}(5\times 10^3)$ and  $\sim {\cal O}(5\times 10^4)$, respectively for very large DM masses.

\begin{figure}
    \centering
    \includegraphics[width=.495\textwidth]{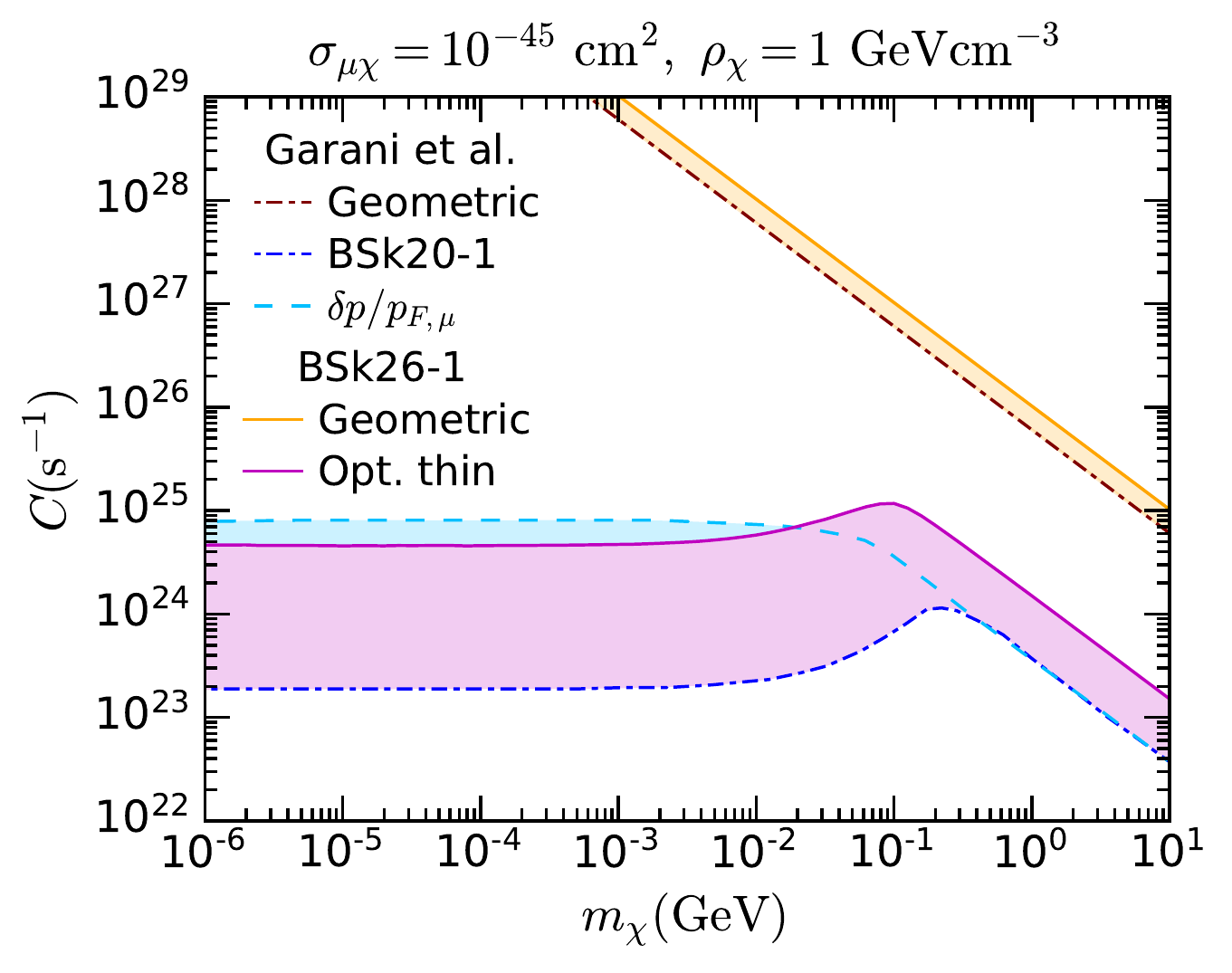}
    \caption{ Capture rate in the optically thin limit for muon targets  (magenta) and geometric (orange) limit as a function of the DM mass for constant cross section $\sigma_{\mu\chi}=10^{-45}\cm^2$,  $\rho_\chi=1\GeV\cm^{-3}$ and BSk26 functional for $\Mstar\simeq1.52\Msun$ and $\Rstar\simeq11.6\km$ denoted as BSk26-1. Capture rate calculations from ref.~\cite{Garani:2018kkd} for a NS configuration with EoS BSk20-1~\cite{Potekhin:2013qqa} equivalent to BSk26-1, are shown for comparison. 
    }
    \label{fig:Cratecomp}
\end{figure}

The DM capture rate for muon targets was calculated in ref.~\cite{Garani:2018kkd}, for constant cross section and light DM, $m_\chi\leq10\GeV$. That calculation accounts for the NS internal structure and Pauli blocking, but neglects general relativity (GR) corrections and assumes that muons are non-relativistic. 
In order to compare our capture rate calculation with that of ref.~\cite{Garani:2018kkd}, 
as in ref.~\cite{Bell:2020jou}, we have selected a NS model that matches that of Fig.~12 of ref.~\cite{Garani:2018kkd},  namely Model A (BSk20-1):  $\Mstar\simeq1.52\Msun$, $\Rstar\simeq11.6\km$. This new benchmark model is denoted as BSk26-1. 
Note that there are no public fits for chemical potentials and particle abundances for BSk20; however,  as mentioned in section~\ref{sec:NSmodels}, BSk26  yields NS configurations that are almost indistinguishable from those obtained with BSk20~\cite{Perot:2019gwl}. 

In Fig.~\ref{fig:Cratecomp}, we compare both capture rate calculations for  $\sigma_{\mu\chi}=10^{-45}\cm^2$ and the same assumptions about  $\rho_\chi$, $v_\star$ and $v_d$ as in ref.~\cite{Garani:2018kkd}. Comparing the geometric limit, Eq.~\ref{eq:capturegeom}  (solid orange), which properly accounts for gravitational focusing in NSs, with the non-relativistic computation in ref.~\cite{Garani:2018kkd} (dot-dashed brown), we observe a $\sim 67 \%$ enhancement, due to the $1/B(\Rstar)$ factor that encodes GR corrections~\cite{Goldman:1989nd,Kouvaris:2007ay}. 
In the region not affected by Pauli blocking, $m_\chi \gtrsim m_\mu$, our  calculation in the optical thin limit (solid magenta) exceeds that of ref.~\cite{Garani:2018kkd} (dot-dashed blue) by a factor of $\sim 4$, which increases as we move to the Pauli suppressed region where our computation is more than one order of magnitude higher. Unlike ref.~\cite{Garani:2018kkd}, our formalism incorporates GR corrections and made use of relativistic kinematics (recall that muons in NSs are mildly relativistic). We also show in dashed light blue, an estimation of the capture rate using the approximation $\delta p/p_{F,\mu}\sim m_\chi v_{esc}/p_{F,\mu}$  for $m_\chi < m_\mu$~\cite{McDermott:2011jp}, where $p_{F,\mu}$ is the muon Fermi momentum and $v_{esc}$ is the escape velocity. This approximation overestimates the capture rate by a factor of approximately 2 in the Pauli blocked region below 10 MeV and underestimates it in the region of larger DM masses.

\subsection{Finite Temperature Effects and Evaporation}
\label{sec:tempcorr}

\begin{figure}[t]
    \centering
    \includegraphics[width=0.5\textwidth]{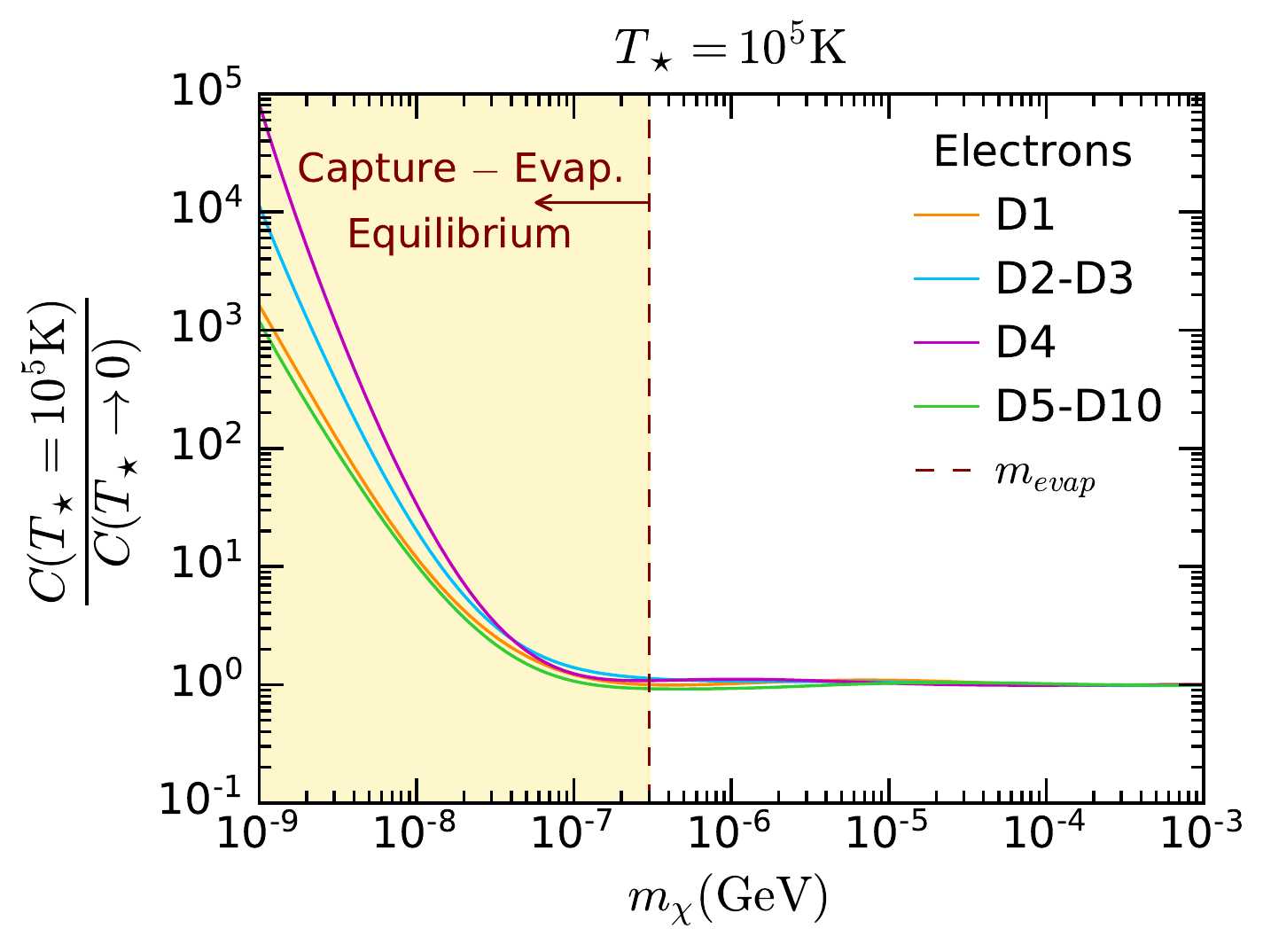}
    \caption{Finite temperature effects on the capture rate for electron targets, assuming the NS model BSk24-2. The DM mass range where capture and evaporation are expected to be in equilibrium is shaded in yellow. The dashed brown line corresponds to the evaporation mass.}
    \label{fig:CfiniteT_e}
\end{figure}

In section~\ref{sec:caprateresults}, we have restricted our computation of the capture rates to the DM mass range $m_\chi\in[1\keV,10^8\GeV]$. It is worth noting that this calculation can also be performed for smaller or larger DM masses. Note, however, that the analytic expressions for the DM interaction rate in ref.~\cite{Bell:2020jou} and appendix~\ref{sec:intrates} were derived in the zero temperature approximation. Therefore, they can be used safely only for $m_\chi\gg \Tstar$, where $\Tstar$ is the NS temperature. 
For $m_\chi\lesssim \mathcal{O}(10)\Tstar$, thermal effects play an important role and increase the capture rate of very light DM \citep{Garani:2018kkd}. Consequently, the complete Fermi Dirac distribution should be used in Eqs.~\ref{eq:responsefunc} and \ref{eq:intrate}. To illustrate the effect of the NS temperature,  we show in Fig.~\ref{fig:CfiniteT_e} the ratio of the capture rate in a NS with 
$\Tstar=10^5\K\simeq8.6\eV$ to the corresponding $C$ rate in the $\Tstar\rightarrow0$ limit, assuming scattering on electrons, the targets for which this effect is most relevant. From this figure, we immediately notice that the ratio starts to depart from  $1$ at $m_\chi\sim100\eV\sim10T_\star$ for all operators. Operators whose matrix element depends on higher powers of the exchanged momentum  $t$ feature a larger increment in the capture rate due to finite temperature. In fact, the operator D4 ($|\overline{M}|^2\propto t^2$) receives the largest correction, followed by D2-D3 (whose $|\overline{M}|^2$ is a linear combination of $t^1,t^2$), D1 ($|\overline{M}|^2$ is a linear combination of $t^0,t^1,t^2$) and finally by  D5-D10 (whose $|\overline{M}|^2$ include all powers of the kind $t^n s^m$).

In the very light DM regime, there is another process we should be aware of: evaporation. This occurs when the dark matter up-scatters to a state where the final DM kinetic energy is larger than the energy required to escape the star, and hence DM particles are expelled. Thus, opposite to capture, evaporation drains energy from the star. 
To estimate the evaporation rate, we convolve the DM distribution within the star, with the interaction rate for up-scattering, $\Gamma_{+}^-$, retaining the temperature dependence.
Assuming the DM distribution to be isothermal with temperature $T_\chi=\Tstar$, we have 
\begin{equation}
n_\chi^{iso}(r,E_\chi) = \frac{n_c}{1+e^{\frac{E_\chi-m_\chi\left(\frac{1}{\sqrt{B(r)}}-1\right)}{\Tstar}}} \simeq 
\frac{ \exp\left[-\frac{E_\chi-m_\chi\left(\frac{1}{\sqrt{B}}-1\right)}{\Tstar}\right]}{4\pi \int_0^{\Rstar} dr r^2  \int_0^{m_\chi\left(\frac{1}{\sqrt{B}}-1\right)} dE_\chi \exp \left[{-\frac{E_\chi-m_\chi\left(\frac{1}{\sqrt{B}}-1\right)}{\Tstar}}\right]}, 
\end{equation}
where $n_c$ is the DM number density at the centre of the star, while the interaction rate for up-scattering, $\Gamma_{+}^-$, is  
\begin{equation}
\dfrac{d\Gamma_{+}^-}{dq_0}\left(q_0,\Tstar\right) = -\frac{e^{q_0/\Tstar}}{1-e^{q_0/\Tstar}} \dfrac{d\Gamma^-}{dq_0}\left(q_0\right),\quad q_0<0,
\label{eq:upscattintrate}
\end{equation}
where $\dfrac{d\Gamma^-}{dq_0}$ is the differential interaction rate in the $\Tstar\rightarrow0$ approximation derived in ref.~\cite{Bell:2020jou} and appendix~\ref{sec:intrates} (for details of the derivation of Eq.~\ref{eq:upscattintrate}, see appendix~\ref{sec:intevaprate}). The evaporation rate then reads
\begin{equation}
E \simeq4\pi \int_0^{\Rstar} dr r^2 \int_0^{m_\chi\left(\frac{1}{\sqrt{B}}-1\right)}dE_\chi n_\chi^{iso}(r,E_\chi) \int_{-\infty}^{-\frac{m_\chi\left(\frac{1}{\sqrt{B}}-1\right)-E_\chi}{\Tstar}} dq_0 \dfrac{d\Gamma_{+}^-}{dq_0}\left(q_0,\Tstar\right).  
\label{eq:evaprate}
\end{equation}
When the DM distribution is concentrated very close to the centre of the star, this expression can be approximated by
\begin{equation}
E \sim  \frac{ m_\chi m_\ell^2\sigma_{\ell\chi}}{4\pi^2} \left(\frac{1}{\sqrt{B(0)}}-1\right)^2  \exp\left[{-\frac{m_\chi}{\Tstar}}\left(\frac{1}{\sqrt{B(0)}}-1\right)\right]. \end{equation}

The rate at which DM particles accumulate in NSs is then given by
\begin{equation}
\dfrac{dN_\chi}{dt} = C- E N_\chi,
\end{equation}
assuming that DM annihilation is negligible. The solution of this equation is 
\begin{equation}
N_\chi(\tstar) = C \, \tstar \left( \frac{ 1-e^{-E \, \tstar}}{E \, \tstar} \right),   
\end{equation}
where $\tstar$ is the age of the NS. The term in brackets quantifies the negative contribution of the evaporation process to the total number of accumulated DM particles. Note that this factor is of order 1 except when $E(m_\chi) \, t_\star \gtrsim {\cal O}(1)$. Therefore, we define the evaporation mass as the DM mass for which the previous relation holds, i.e.  $E(m_{evap}) t_\star \sim 1$. 
For DM masses below this threshold, $m_\chi \lesssim m_{evap}$, the capture and evaporation processes are in equilibrium with each other. In that limit, the net energy exchange in the star due to combined effects of DM capture and evaporation would be negligible, and hence we would be unable to constrain DM interactions using the NS temperature as a probe.

Using Eq.~\ref{eq:evaprate}, we find the evaporation mass to be of order $m_{evap}\sim\mathcal{O}(100\Tstar)$ for all scattering targets in old NSs with $\tstar\sim{\cal O}(10 \Gyr)$. For instance, for $\Tstar=10^5\K$ and electron targets, we obtain $m_{evap}\simeq 300\eV$. 
From Fig.~\ref{fig:CfiniteT_e}, we note that the evaporation mass (dashed brown line) is larger than the mass at which temperature effects on the capture rate are important. In other words, evaporation comes into play before finite temperature effects become relevant so that these effects can be  safely neglected when calculating the capture rate with the aim to constrain DM interactions.

\subsection{Threshold Cross Section}
\label{sec:thxs}
In ref.~\cite{Bell:2020jou}, we defined the threshold cross section, $\sigmathl$, as the cross section for which the capture rate   $C(\sigma(\Lambda),m_\chi)$, calculated in the optically thin regime (i.e. without the optical factor $\eta$), is $\sim C_{geom}$. This definition is general and is applicable to both relativistic and non-relativistic targets. 
The threshold cross section restricts the NS sensitivity to DM-target  interactions, since for $\sigma\ge\sigmathl$ the capture rate saturates to the geometric limit $C_{geom}$.

\begin{figure}
\centering
\includegraphics[width=0.49\textwidth]{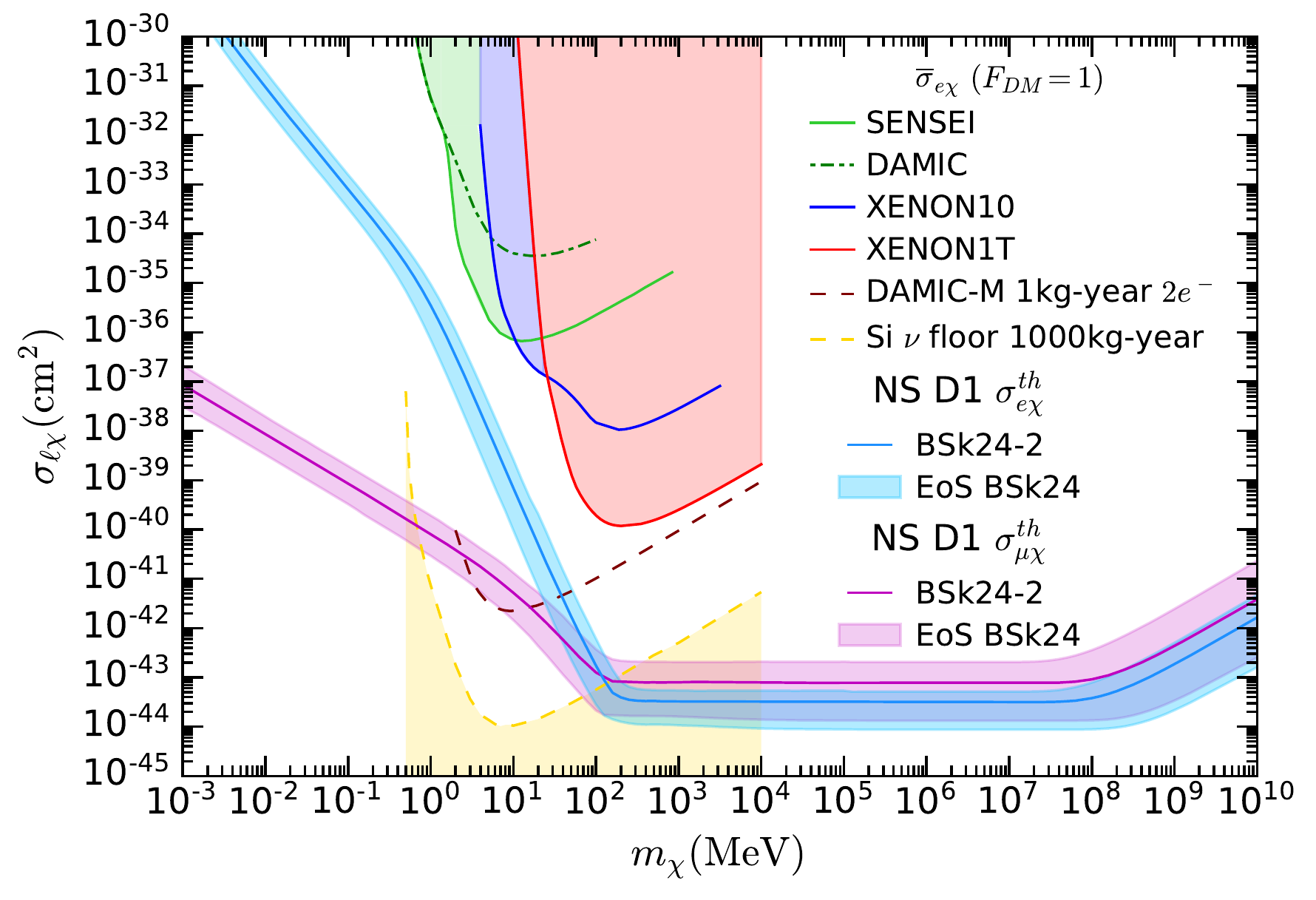}
\includegraphics[width=0.49\textwidth]{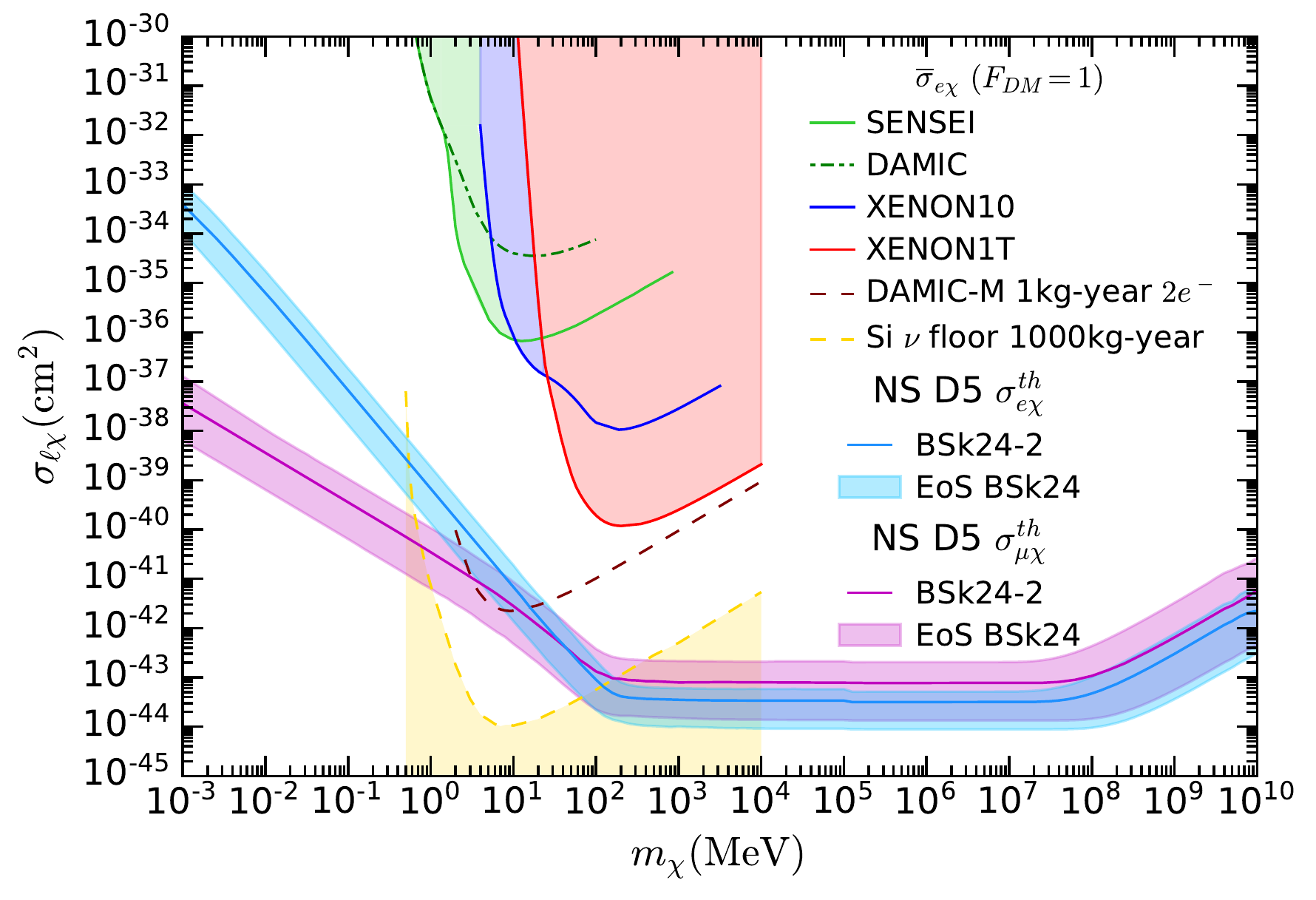}
\caption{DM-lepton threshold cross section for operators D1 (left) and D5 (right) for the EoS BSk24. The solid blue (electron) and magenta (muon) lines represent $\sigmath$, computed assuming the NS model BSk24-2, while the shaded bands represent the expected range due to variation of the EoS. For comparison we show leading electron recoil bounds for heavy mediators from SENSEI~\cite{Barak:2020fql}, DAMIC~\cite{Aguilar-Arevalo:2019wdi}, Xenon10~\cite{Essig:2017kqs}, Xenon1T~\cite{Aprile:2019xxb}, projected sensitivities from DAMIC-M~\cite{Essig:2015cda} as well as the neutrino floor for silicon detectors~\cite{Essig:2018tss}. }
    \label{fig:sigmathe}
\end{figure}

In Fig. \ref{fig:sigmathe}, we  show the threshold cross sections for lepton targets, electrons and muons, and compare them with existing limits and expected sensitivities of future experiments. The neutrino floor for electron recoil experiments for silicon targets~\cite{Essig:2018tss} is shown as a shaded yellow region.
The solid light blue and magenta lines correspond to the value of $\sigma_{th}$ for electrons and muons respectively, calculated using the NS model BSk24-2  ($1.5 M_\odot$), while the shaded bands in light blue and magenta denote the expected range for $\sigma_{th}$ for the two different targets, obtained by varying the NS configuration along the BSk24 family. 
 BSk24-1 ($1M_\odot$) gives the upper bound on $\sigmath$ and BSk24-4 ($2.16 M_\odot$) the lower bound. 
Note that the variation in $\sigmath$ due to the NS EoS increases with the DM mass and for muons goes from about one order of magnitude in the low mas range to two orders of magnitude in the multiple scattering region. For electrons, this effect is slightly less pronounced. 
 All the limits for existing experiments are orders of magnitude weaker than the expected NS reach, with only the future DAMIC-M \citep{Essig:2015cda} (dashed brown line) expected to overcome NS electron scattering sensitivity and approach that of muons, in the DM mass range  $3\MeV\lesssim m_\chi\lesssim 30\MeV$. Moreover, NS sensitivity to DM interactions with lepton targets is expected to be well below the neutrino floor for $m_\chi\gtrsim 100\MeV$ and, in the case of muons, even for $m_\chi\lesssim 1\MeV$. 
Note that NSs have a better sensitivity to vector-vector interactions (operator D5, see right panel) than scalar-scalar interactions (operator D1, see left panel) in the low DM regime for both leptonic targets, especially for electrons, since as mentioned in section~\ref{sec:caprateresults} there is an additional suppression in the capture rate of scalar operators that stems from a $m_e^2 \, t$ term in their scattering amplitudes. 
Similar threshold cross sections can be estimated for the remaining operators.  Operators with s-dependent matrix elements (D6-D10) have  $\sigmath$ that behaves like that of D5 for both electrons and muons. D2 presents the same features as D1 in the sub-GeV regime for electrons, due to the similar shape of their capture rates (see Fig.~\ref{fig:capratesD1D4}) and D3-D4 show a steeper slope in the $m_\chi\lesssim m_e$ region with respect to D1-D2, due to the capture rate dependence on higher powers of $t$  (see Table~\ref{tab:operatorshe} and Fig.~\ref{fig:capratesD1D4}).

\begin{figure}
    \centering
    \includegraphics[width=\textwidth]{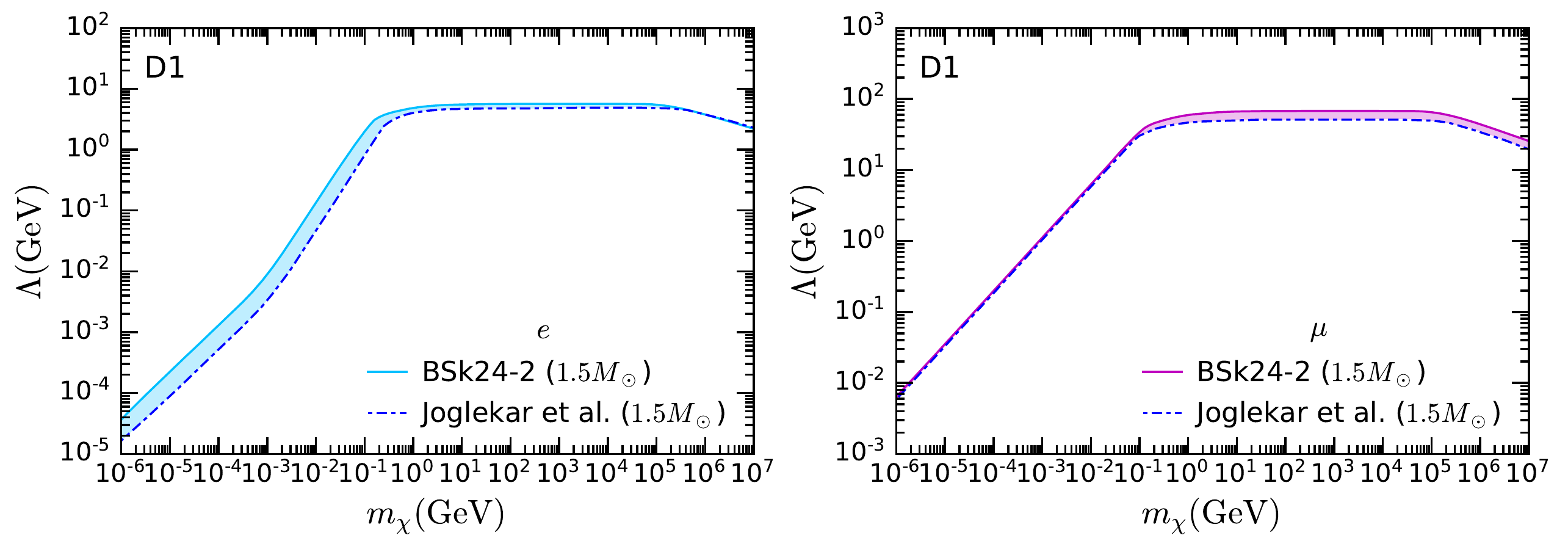}\\
    \vspace*{-0.5em}    
    \includegraphics[width=\textwidth]{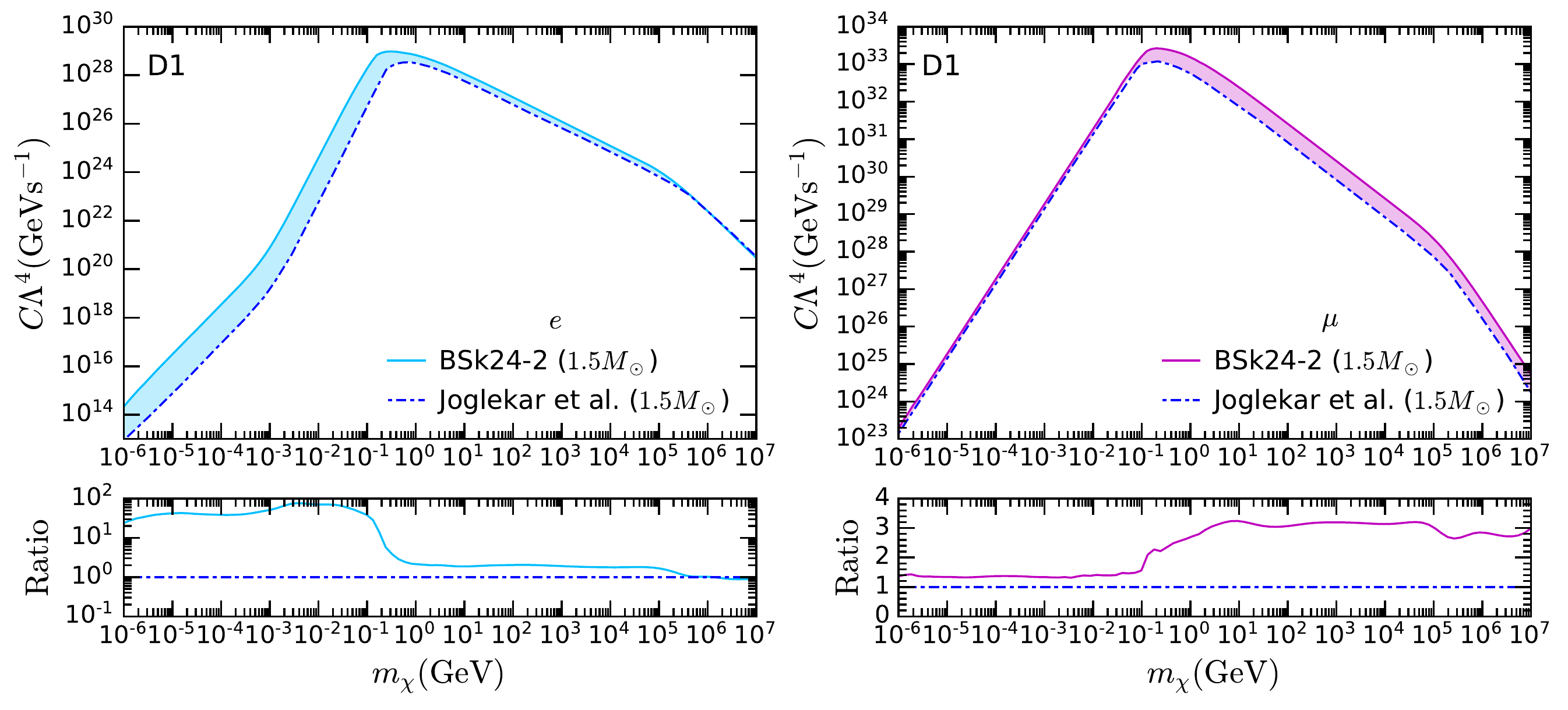}     
    \caption{Comparison of the reach in $\Lambda$ for D1 with the approach of  ref.~\cite{Joglekar:2020liw}. In these references, a NS with constant chemical potentials and particles abundances averaged over the core volume was assumed, these quantities were taken from ref.~\cite{Bell:2019pyc} and are consistent with a NS with EoS BSk24-2. The shaded regions denote the difference in $\Lambda$ (top) and $C\Lambda^4$ (middle) between the two approaches and their ratio is shown in the bottom panels.  }
    \label{fig:D1_Lambda_mdm}
\end{figure}

\begin{figure}
    \centering
    \includegraphics[width=\textwidth]{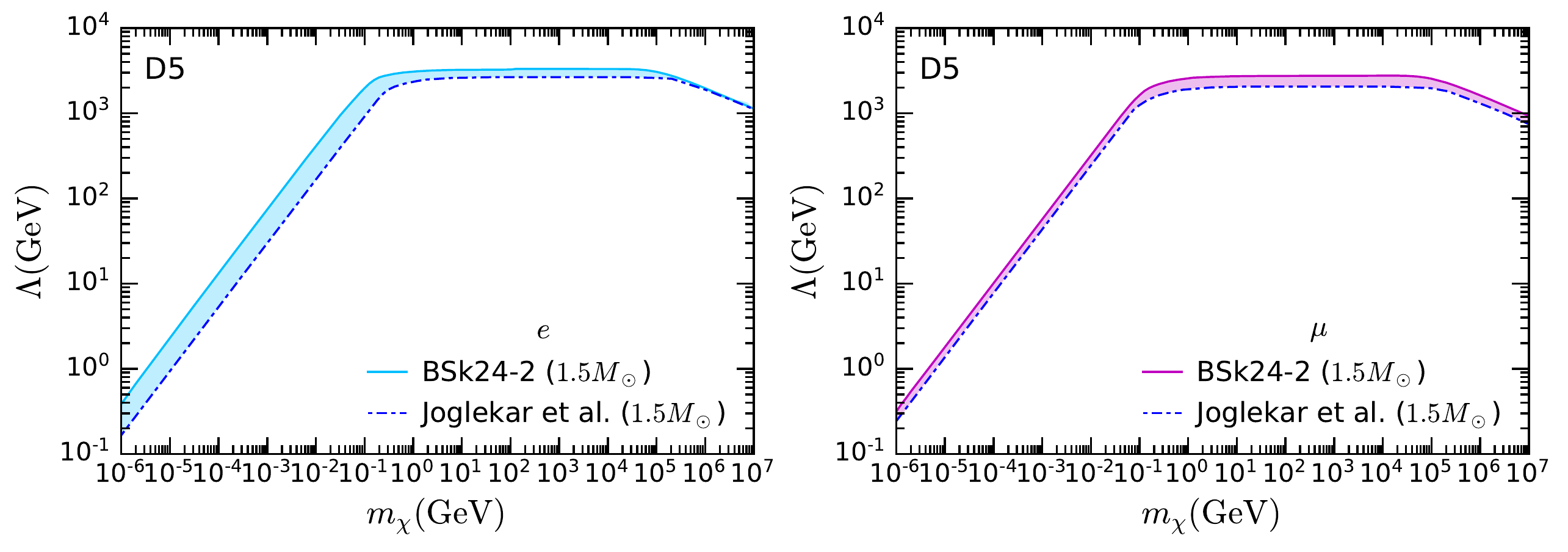}\\
    \vspace*{-0.5em}
    \includegraphics[width=\textwidth]{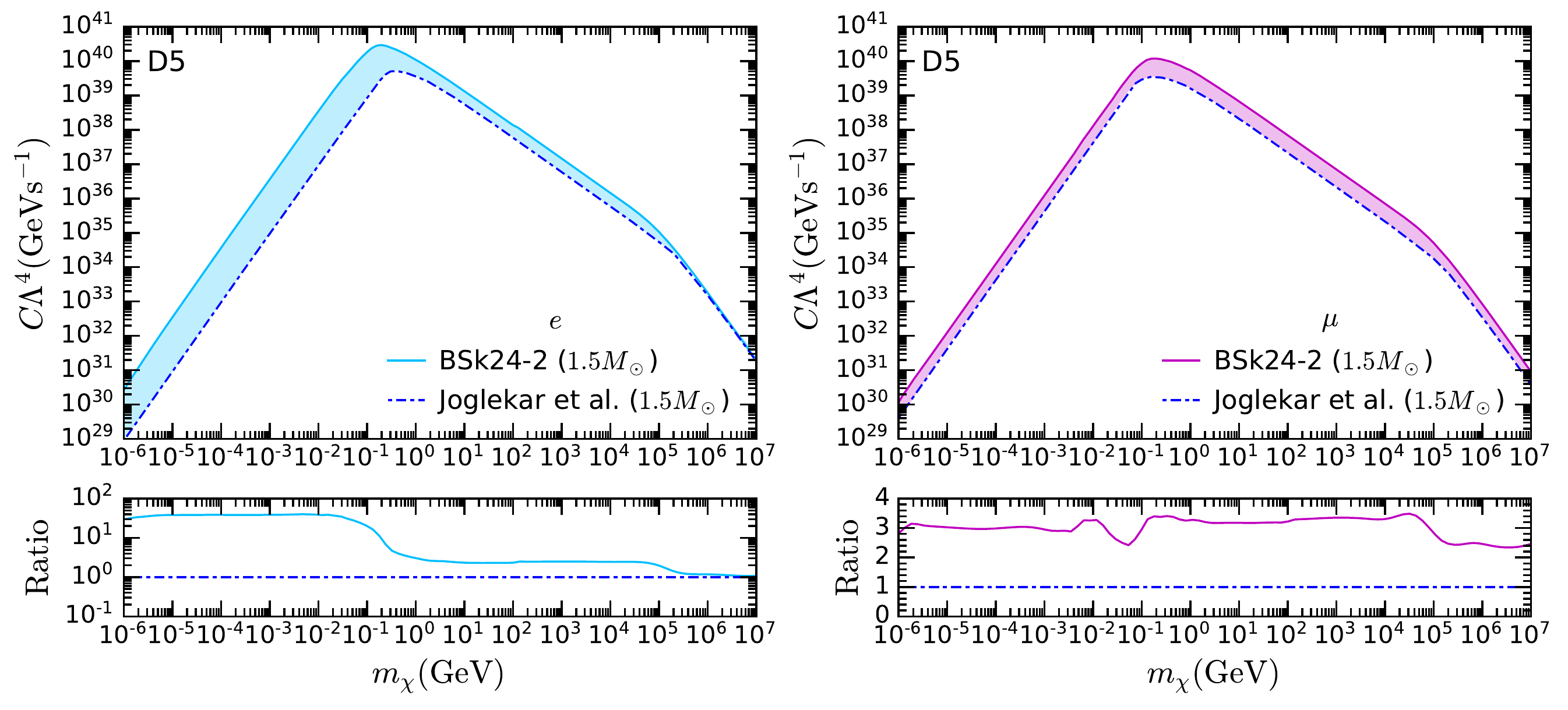}    
    \caption{Comparison of the reach in $\Lambda$ for D5 with the approach of  refs.~\cite{Joglekar:2019vzy,Joglekar:2020liw}. The shaded regions denote the difference in $\Lambda$ (top) and $C\Lambda^4$ (middle) between the two approaches and their ratio is shown in the bottom panels. }
    \label{fig:D5_Lambda_mdm}
\end{figure}

In Figs.~\ref{fig:D1_Lambda_mdm} and ~\ref{fig:D5_Lambda_mdm}, we compare our results for D1 and D5 with those of refs.~\cite{Joglekar:2019vzy,Joglekar:2020liw}~\footnote{Note that the Yukawa couplings for scalar and pseudoscalar operators in refs.~\cite{Joglekar:2019vzy,Joglekar:2020liw} are embedded into the cutoff scale $\Lambda$.}. The formalism in refs.~\cite{Joglekar:2019vzy,Joglekar:2020liw} is valid for relativistic and non-relativistic targets in a broad mass range, but neglects the DM velocity distribution and the NS internal structure. Instead, constant chemical potentials and particle abundances, averaged over the core volume, are assumed. These quantities correspond to the NS model BSk24-2 and were calculated in ref.~\cite{Bell:2019pyc}. In the top panels, we compare the reach in $\Lambda$ for DM-lepton scattering cross sections in refs.~\cite{Joglekar:2019vzy,Joglekar:2020liw} with the cutoff scale we obtain for the maximum capture rate  $C(\Lambda,m_\chi)=C_{geom}$. 
Our results differ the most for electron targets in the Pauli suppressed region by a factor of $\sim 2.5$ and we find Pauli blocking is active at a slightly lighter DM mass. Recall that we have obtained the radial profiles for chemical potentials and number densities with a unified EoS, i.e. for the core and the crust, and note that most light DM particles whose interactions are subject to Pauli blocking are captured close to the surface~\cite{Bell:2020jou}. The difference between our approach and that of refs.~\cite{Joglekar:2019vzy,Joglekar:2020liw} is reduced to a factor of  $\sim 1.25$ in the intermediate mass region and there is almost no difference in the large mass regime, except for the DM mass at which multiple scattering becomes relevant, which in our case is once again slightly lighter. For muons, we find a $\Lambda$ that is, on average, a factor $\sim 1.33$ (D5) greater than that of refs.~\cite{Joglekar:2019vzy,Joglekar:2020liw} along the whole DM mass range and is in almost perfect agreement in the $m_\chi\lesssim m_\mu$ region for D1.

In the middle panels of Figs.~\ref{fig:D1_Lambda_mdm} and ~\ref{fig:D5_Lambda_mdm}, for operators D1 and D5 respectively, we show how these apparently small differences in the two approaches translate to differences in the capture rate. To that end, we compare $C \Lambda^4 = \Lambda^4 C_{geom}$  obtained with the two formalisms. Since the geometric limit of the capture rate is not defined in refs.~\cite{Joglekar:2019vzy,Joglekar:2020liw}, we use a definition similar to Eq.~\ref{eq:capturegeom} and compliant with assumptions made by these authors. For electron scattering, we see that the formalism that does not account for the NS internal structure underestimates the capture rate in the region affected by Pauli blocking by a factor $\sim40$ (bottom LH panels of Figs.~\ref{fig:D1_Lambda_mdm} and ~\ref{fig:D5_Lambda_mdm}). This difference is slightly larger in the region where the Pauli suppression is stronger, becoming almost a factor of $\sim 100$ for D1 in the range $m_e\lesssim m_\chi \lesssim 100\MeV$. For muons, the difference between the two approaches is less pronounced, with a maximum ratio of $\sim 3.5$ for both operators.

\section{Conclusions}
\label{sec:conclusions}

Neutron stars (NSs) are potential cosmic laboratories to study dark matter (DM) interactions with ordinary matter under extreme conditions. Gravitational focusing enhances the rate at which DM particles can accumulate in these stars. Thus, NSs emerge as potential DM probes,  complementary to direct detection experiments which are restricted  by recoil thresholds  and small momentum transfers. DM scattering off NS targets is, however, not free of limitations; in the sub-GeV regime DM scattering off strongly degenerate targets is suppressed by Pauli blocking and in the large mass region multiple collisions are required to capture heavy DM. Moreover, there is a natural threshold for the maximum cross section that can be probed in NSs, above which the capture rate saturates to its geometric limit. 

NSs are systems in beta equilibrium such that, even though they 
are primarily composed of degenerate neutrons, protons and electrons are present throughout the star with abundances of order a few percent.  Muons are also found in NS cores at higher densities. Unlike nucleons, the leptonic species in the NS are relativistic.
In this paper, we have examined the reach of NSs to probe the interactions of fermionic DM with the leptonic NS constituents, in the context of an effective field theory (EFT).  
To that end, we generalised our formalism to calculate the DM interaction rate, presented in ref.~\cite{Bell:2020jou}, to enable it to handle
any differential cross section parametrized in terms of the Mandelstam variables $s$ and $t$. With this extended formalism, we calculated the capture rate for the full list of dimension 6 effective operators for a broad DM mass range, properly including Pauli blocking, multiple scattering, NS internal structure and general relativity (GR) corrections. 

To be consistent, the aforementioned calculation requires knowledge of the microscopic properties of the target species, such as chemical potential, number density and abundance, as well as GR corrections.  These quantities have a radial dependence and hence require the assumption of a NS equation of state (EoS). This is particularly relevant for leptons, as their particle fractions are heavily dependent on the NS mass. 
To account for that uncertainty, we  have assumed the unified EoS with Brussels-Montreal functional BSk24, which is well motivated by observations.

We find that scattering off muons dominates the leptonic contribution to the capture rate for scalar and pseudoscalar DM-lepton interactions, despite the muon abundance being lower than that of electrons. This is due to the fact that the couplings for these interactions scale with the lepton mass. For other interaction types, electrons and muons have the same coupling scale. In spite of that, and the fact that electrons are ultra-relativistic while muons are only mildly relativistic, the capture rates for electron and muon targets are comparable, a consequence of the large muon mass and lower muon chemical potential. This effect is enhanced in heavy NSs where muon and electron number densities are very similar. 

The NS sensitivity to DM interactions with leptons greatly surpasses that of any current direct detection (DD) experiment, particularly in the sub-GeV DM regime. Only future DD  experiments such as DAMIC-M could be competitive and, even then, only in a narrow mass range $3\MeV \lesssim m_\chi\lesssim30\MeV$. Finally, note that the evaporation mass, the DM mass below which capture and evaporation processes are expected to be in equilibrium, is much smaller for NSs than for other stars or planets. This provides sensitivity to sub-MeV DM with scattering cross sections even below the DD neutrino floor. These findings are particularly relevant for leptophilic DM for which couplings to nucleons arise at loop level.

\section*{Acknowledgements}
NFB and SR were supported by the Australian Research Council and MV by the Commonwealth of Australia. We thank Filippo Anzuini and Tony Thomas for helpful discussions.

\appendix

\section{Interaction rate for s-dependent amplitudes}
\label{sec:intrates}

In ref.~\cite{Bell:2020jou}, we obtained analytic expressions for the DM interaction rate, for squared matrix elements that depend on $t$,  but not on the centre of mass energy $s$. 
In the following, we generalise our previous result to  $|\overline{M}|^2$ that can be written as a polynomial function of the variables $s$ and $t$, i.e. $ |\overline{M}|^2 = \alpha s^m t^n$, where $n$ are $m$ are integers and $\alpha$ is a constant.

Following refs.~\citep{Bertoni:2013bsa} and \cite{Bell:2020jou}, we define the  DM scattering rate as
\begin{eqnarray}
\Gamma &=& \int \frac{d^3k^{'}}{(2\pi)^3} \frac{1}{(2E_\chi)(2E^{'}_\chi)(2m_i)(2m_i)}\Theta(E^{'}_\chi-m_\chi)\Theta(q_0)S(q_0,q), \label{eq:intratedef}\\
S(q_0,q) &=& 2\int \frac{d^3p}{(2\pi)^3}\int \frac{d^3p^{'}}{(2\pi)^3} \frac{m_i^2}{E_i E^{'}_i}|\overline{M}|^2 (2\pi)^4\delta^4\left(k_\mu+p_\mu-k_\mu^{'}-p_\mu^{'}\right)\nonumber\\
 &&\times\fFD(E_i)(1-\fFD(E^{'}_i))\Theta(E_i-m_i)\Theta(E^{'}_i-m_i),
\end{eqnarray}
where $k^\mu=(E_\chi$,$\vec{k})$, $k^{'\mu}=(E^{'}_\chi,\vec{k'})$ are the DM initial and final momenta, $p^\mu=(E_i,\vec{p})$ and
$p^{'\mu}=(E^{'}_i,\vec{p'})$ are the target particle initial and final momenta, and  $q_0=E_i^{'}-E_i$ is the DM energy loss. Note that now $\Msq$ depends on $p$, so we leave it inside the response function $S(q_0,q)$.

Integrating the response function over $d^3p^{'}$ using the delta function leaves
\begin{equation}
S(q_0,q) = \frac{1}{2\pi^2}\int d^3p \frac{m_i^2}{E_i E^{'}_i} |\overline{M}|^2 \delta\left(q_0+E_i-E^{'}_i\right)\fFD(E_i)(1-\fFD(E^{'}_i))\Theta(E_i-m_i)\Theta(E^{'}_i-m_i).      
\end{equation}
After that the final target energy is fixed to
\begin{equation}
E^{'}_i(E_i,q,\theta) = \sqrt{m_i^2+(\vec{p}+\vec{q})^2} = \sqrt{E_i^2+q^2+2qp\cos\theta} > m_i, \quad\forall p,q,\theta, |\cos\theta|<1,
\end{equation} 
where $\theta$ is the angle between $\vec{p}$ and $\vec{q}$. To perform the integral  over $d^3p$ we change it to $d^3p = pE_i\,dE_i\,d\cos\theta d\phi$ and use the delta function to integrate over $\theta$~\citep{Bertoni:2013bsa,Reddy:1997yr}. Note that this gives rise to $\Theta(1-\cos^2\theta)$~\citep{Reddy:1997yr}. 
Then, we obtain
\begin{eqnarray}
S(q_0,q) = \alpha t^n \frac{m_i^2}{2\pi^2 q}\int dE_i d\phi s^m  \fFD(E_i)(1-\fFD(E_i+q_0))\Theta(E_i)  \Theta(1-\cos^2\theta(q,q_0,E_i)).
\end{eqnarray}
Using 
\begin{equation}
\cos\theta(q,q_0,E_i) =  \frac{q_0^2-q^2+2E_iq_0}{2q\sqrt{E_i^2-m_i^2}},
\end{equation}
we can determine the integration interval for $E_i$. 
For the $q^2>q_0^2$ case, $q^\mu$ is expected to be space-like, $t=q_\mu q^\mu <0$, and the response function becomes
\begin{eqnarray}
S^{-}(q_0,q) = \alpha t^n \frac{m_i^2}{2\pi^2 q}\int_{E_i^{\, t^{-}}}^\infty dE_i  \fFD(E_i)(1-\fFD(E_i+q_0))\int d\phi \, s^m,
\end{eqnarray}
where $E_i^{t^-}$ is the minimum energy of the neutron before the collision, which is obtained from kinematics and given by 
\begin{equation}
E_i^{\, t^{-}} = -\left(m_i+\frac{q_0}{2}\right) + \sqrt{\left(m_i+\frac{q_0}{2}\right)^2+\left(\frac{\sqrt{q^2-q_0^2}}{2}-\frac{m_i q_0}{\sqrt{q^2-q_0^2}}\right)^2}. 
\end{equation} 
To  perform the integration over the azimuth angle $\phi$ we rewrite $s$ in terms of the other kinematic variables, $q,q_0,E_i,E_\chi$, $s= m_i^2+m_\chi^2 + 2 E_\chi E_i - 2\vec{p}\cdot \vec{k}$, 
where 
the value of the scalar product of the two momenta is
\begin{eqnarray}
    \vec{p}\cdot \vec{k} &=& \frac{\left(q^2-q_0^2+2E_\chi q_0\right)\left(q_0^2-q^2+2q_0E_i\right)}{4q^2}  \nonumber \\ &+&  \sqrt{E_\chi^2-m_\chi^2-\frac{\left(q^2-q_0^2+2E_\chi q_0\right)^2}{4q^2}}\sqrt{E_i^2-m_i^2-\frac{\left(q^2-q_0^2-2E_iq_0\right)^2}{4q^2}}\cos\phi. \label{eq:scalprodexpr}
\end{eqnarray}
We are mostly interested in values of $m=1,2$. For instance, for $m=1$ 
\begin{eqnarray}
S^-(q_0,q) &=& \alpha t^n \frac{m_i^2}{2\pi^2 q}\int_{E_i^{\, t^{-}}}^\infty dE_i  \fFD(E_i)(1-\fFD(E_i+q_0)) \int d\phi \, s\\
&=& \alpha t^n \frac{m_i^2}{\pi q}\int_{E_i^{\, t^{-}}}^\infty dE_i  \fFD(E_i)(1-\fFD(E_i+q_0))\nn\\
&\times&\left(m_\chi^2+m_i^2+2E_\chi E_i-2\frac{\left(q^2-q_0^2+2E_\chi q_0\right)\left(q^2-q_0^2-2q_0E_i\right)}{4q^2}\right)\label{eq:respionses1}
\end{eqnarray}
From now on, we do not give explicit expressions for the integrals, due to their length, but just sketch the procedure to easily  obtain the solutions using any symbolic calculation language. 
The previous expression \ref{eq:respionses1} can be written as
\begin{eqnarray}
S^-(q_0,q) &=& \alpha t^n \frac{m_i^2}{\pi q}\int_{E_i^{\, t^{-}}}^\infty dE_i  \fFD(E_i)(1-\fFD(E_i+q_0)) \frac{\mathcal{U}_m(q^2,q_0,E_\chi,E_i)}{q^{2m}}, 
\end{eqnarray}
where $\mathcal{U}_m(q^2,q_0,E_\chi,E_i)$ is a polynomial of degree $m$ in $E_i$ that can be rewritten as 
\begin{eqnarray}
    \mathcal{U}_1(q^2,q_0,E_\chi,E_i) &=& \mathcal{V}_{1,0}(q^2,q_0,E_\chi) + \mathcal{V}_{1,1}(q^2,q_0,E_\chi) E_i,\\
    \mathcal{U}_2(q^2,q_0,E_\chi,E_i) &=& \mathcal{V}_{2,0}(q^2,q_0,E_\chi) + \mathcal{V}_{2,1}(q^2,q_0,E_\chi) E_i + \mathcal{V}_{2,2}(q^2,q_0,E_\chi) E_i^2, 
\end{eqnarray}
where $\mathcal{V}_{m,i}(q^2,q_0,E_\chi)$ are also polynomials. 
We therefore need to calculate integrals of the form
\begin{equation}
    \int dE_i E_i^j  \fFD(E_i)(1-\fFD(E_i+q_0)), \qquad 0\le j \le m.
    \label{eq:intEn}
\end{equation}
Using
\begin{eqnarray}
&&1-\fFD(E_i+q_0) = \fFD(-E_i-q_0), \\
&&F_0(x,z) = \int dx \fFD(x)\fFD(-x-z) = \frac{e^z \left[\log \left(e^{x+z}+1\right)-\log \left(e^x+1\right)\right]}{e^z-1},
\end{eqnarray}
we solve the integrals, e.g. for $m=1$
\begin{eqnarray}
 \int dE_i E_i  \fFD(E_i)(1-\fFD(E_i+q_0)) 
 &=& \int dE_i \left(E_i-\muFi\right)  \fFD(E_i)(1-\fFD(E_i+q_0)) \nonumber \\
 &&+ \muFi\int dE_i  \fFD(E_i)(1-\fFD(E_i+q_0)) \nn\\
 &=& F_1\left(E_i-\muFi,q_0\right) + \muFi F_0\left(E_i-\muFi,q_0\right),
\end{eqnarray}
where $F_1(x,z)$ is obtained by integrating by parts
\begin{eqnarray}
F_1(x,z) &=& \int dx x \fFD(x)\fFD(-x-z) \nn\\
&=&\frac{e^z \left[x\left(\log \left(e^{x+z}+1\right)-\log \left(e^x+1\right)\right)+PL\left(2,-e^{x+z}\right)-PL\left(2,-e^{x}\right)\right]}{e^z-1}, 
\end{eqnarray}
and $PL$ is the PolyLog function. Similarly, we can solve the $m=2$ integral over $E_i$ for 
\begin{equation}
F_2(x,z) = \int dx x^2 \fFD(x)\fFD(-x-z).     
\end{equation}

For $F_0$,  three distinct regimes can be identified as noted in ref.~\cite{Bell:2020jou} ($s$-independent case)
\begin{eqnarray}
&E_i&>\muFi,\\
\muFi-q_0<&E_i&<\muFi,\\
&E_i&<\muFi -q_0.
\end{eqnarray}
Next, we use the following results for each of the above $E_i$ intervals, which are valid in the $\Tstar\rightarrow0$ limit
\begin{eqnarray}
\lim_{\Tstar\rightarrow0} \Tstar F_0(E_i/\Tstar,q_0/\Tstar) &=& q_0, \quad E_i>\muFi,\\
\lim_{\Tstar\rightarrow0} \Tstar F_0(E_i/\Tstar,q_0/\Tstar) &=& E_i+q_0-\muFi, \quad \muFi-q_0<E_i<\muFi,\\
\lim_{\Tstar\rightarrow0} \Tstar F_0(E_i/\Tstar,q_0/\Tstar) &=& 0, \quad E_i<\muFi-q_0. 
\end{eqnarray}
The expressions above, which resemble a step function with a smooth transition, can be recast in terms of the function
\begin{align}
g_0(x) =  \begin{cases}
\, 1 \quad &x>0, \\
\, 1+x \quad &-1<x<0,\\
\, 0 \quad &x<-1.
\end{cases}
\end{align} 
such that
\begin{eqnarray}
\lim_{\Tstar\rightarrow0} \Tstar F_0(E_i/\Tstar,q_0/\Tstar) &=& q_0 \,  g_0\left(\frac{E_i-\muFi}{q_0}\right).
\end{eqnarray}
Then, the contribution of $F_0$ to the response function in the $\Tstar\rightarrow0$ limit is
\begin{eqnarray}
S^{-}_{0}(q_0,q) &=&  \alpha t^n\frac{m_i^2q_0}{\pi q}\left[1-g_0\left(\frac{E_i^{\,t^{-}}-\muFi}{q_0}\right)\right]=\alpha t^n\frac{m_i^2q_0}{\pi q}h_0\left(\frac{E_i^{\,t^{-}}-\muFi}{q_0}\right),
\label{eq:Sminust}
\end{eqnarray}
where 
\begin{align}
h_0(x) =  1-g_0(x) =\begin{cases}
\, 0 \quad &x>0, \\
\, -x \quad &-1<x<0,\\
\, 1 \quad &x<-1. 
\end{cases}
\end{align} 
Note that Eq.~\ref{eq:Sminust} is the result we found in ref.~\cite{Bell:2020jou}. 

 We proceed in a similar way for $F_1, F_2$ and obtain
\begin{eqnarray}
\lim_{\Tstar\rightarrow0} \Tstar^2 F_1(E_i/\Tstar,q_0/\Tstar) &=& -\frac{q_0^2}{2} \,  g_1\left(\frac{E_i-\muFi}{q_0}\right),\\
\lim_{\Tstar\rightarrow0} \Tstar^3 F_2(E_i/\Tstar,q_0/\Tstar) &=& \frac{q_0^3}{3} \,  g_2\left(\frac{E_i-\muFi}{q_0}\right),
\end{eqnarray}
where
\begin{align}
g_1(x) =  \begin{cases}
\, 1 \ &x>0, \\
\, 1-x^2 \ &-1<x<0,\\
\, 0 \ &x<-1, 
\end{cases}
\qquad
g_2(x) =  \begin{cases}
\, 1 \ &x>0, \\
\, 1+x^3 \ &-1<x<0,\\
\, 0 \ &x<-1.
\end{cases}
\end{align} 
Then, the general expression for the response function is
\begin{eqnarray}
S^-_{m}(q_0,q) &=& \alpha t^n \frac{m_i^2}{\pi q^{1+2m}} \sum_{i=0}^m\mathcal{V}_{m,i}(q^2,q_0,E_\chi) \frac{q_0^{1+i}}{1+i} h_i\left(\frac{E_i^{\,t^{-}}-\muFi}{q_0}\right),
\label{eq:Sminus}
\end{eqnarray}
where
\begin{align}
h_1(x) = 1-g_1(x) = \begin{cases}
\, 0 \ &x>0, \\
\, x^2 \ &-1<x<0,\\
\, 1 \ &x<-1,
\end{cases}
\qquad
h_2(x) = 1-g_2(x) = \begin{cases}
\, 0 \ &x>0, \\
\, -x^3 \ &-1<x<0,\\
\, 1 \ &x<-1.
\end{cases}
\end{align} 
Comparing to the case where the amplitude does not depend on $s$, there are two additional transition functions.

We now return to the scattering rate 
\begin{eqnarray}
\Gamma^- 
&=& \alpha \int \frac{d\cos\theta k^{'2}dk^{'}}{64\pi^2E_\chi E^{'}_\chi m_i^2} t^n \Theta(E_\chi-q_0-m_\chi)\Theta(q_0)S^-(q_0,q), 
\end{eqnarray}
change variables from $k^{'},\cos\theta$ to $q_0,q$, 
\begin{eqnarray}
q_0 &=& E_\chi-\sqrt{k^{'2}+m_\chi^2},\label{eq:q0}\\
q^2 &=& k^2+k^{'2}-2k k^{'}\cos\theta, \label{eq:q}
\end{eqnarray}
and substitute in the result for $S^-$, Eq.~\ref{eq:Sminus}, to obtain, 
\begin{equation}
\Gamma^{-} 
= \frac{\alpha}{64\pi^3E_\chi k }\int dq dq_0 \frac{q_0 t^n}{q^{1+2m}} \sum_{i=0}^m \mathcal{V}_{m,i}(q^2,q_0,E_\chi) \frac{q_0^{i}}{1+i} h_i\left(\frac{E_i^{\, t^{-}}-\muFi}{q_0}\right) \Theta(E_\chi-q_0-m_\chi)\Theta(q_0). 
\end{equation}
To simplify the integration over $t$, we define $t_E=-t=q^2-q_0^2$, which leads to 
\begin{equation}
\Gamma^{-}=\frac{(-1)^n \alpha}{2^7\pi^3E_\chi k }\sum_{i=0}^m\int_0^{E_\chi-m_\chi}\frac{q_0^{i+1} dq_0}{i+1} \int \frac{t_E^n dt_E }{\left(q_0^2+t_E\right)^{\frac{1}{2}+m}} \mathcal{V}_{m,i}(t_E+q_0^2,q_0,E_\chi) h_i\left(\frac{E_i^{\,t^{-}}-\muFi}{q_0}\right).
\end{equation}

As $\mathcal{V}_{m,i}(t_E+q_0^2,q_0,E_\chi)$ are polynomials, we need to calculate integrals of the form
\begin{equation}
    \int_0^{E_\chi-m_\chi}q_0^{l+1} dq_0 \int \frac{t_E^j dt_E }{\left(q_0^2+t_E\right)^{\frac{1}{2}+m}} h_i\left(\frac{E_i^{\,t^{-}}-\muFi}{q_0}\right),
\end{equation}
which can be solved by decomposing the integration interval using the primitives of
\begin{eqnarray}
   \tilde{f}_1(t_E,q_0) =  \frac{t_E^j  }{\left(q_0^2+t_E\right)^{\frac{1}{2}+m}},\qquad 
    \tilde{f}_2(t_E,q_0) = \frac{t_E^j  }{\left(q_0^2+t_E\right)^{\frac{1}{2}+m}} \left(\frac{E_i^{\,t^{-}}-\muFi}{q_0}\right)^{k+1}.
\end{eqnarray}
Using the operator defined in ref.~\cite{Bell:2020jou} that encodes the $t_E$ integral over the correct intervals, 
\begin{align}
    \mathcal{I}(\tilde{f}(t),t_1^+,t_2^+,t_1^-,t_2^-) =& \sum_{k=1,2}\sum_{j=1,2} \left(\tilde{F}(t_k^+)-\tilde{F}(t_j^-)\right)\Theta\left(t_{3-k}^+-t_k^+\right)\Theta\left(t_k^+-t_j^-\right)\nonumber\\
    &\times\Theta\left(t_j^--t_{3-j}^-\right),\\
    \tilde{F}(t) =& \int dt \, \tilde{f}(t), 
\end{align}
we obtain $\Gamma^{-}$ as a linear combination of terms of the kind
\begin{align}
\Gamma^{-}(E_\chi) \propto \,& \left[\int_0^{E_\chi-m_\chi}q_0^{k+1} dq_0 \,\, \mathcal{I}\left(\tilde{f}_1(t_E,q_0),t_E^+,t_{\mu^-}^+,t_E^-,t_{\mu^-}^-\right)\Theta(\muFi-q_0)\right. \nonumber\\
&+ \int_0^{E_\chi-m_\chi} q_0^{k+1} dq_0 \,\, \mathcal{I}\left(\tilde{f}_2(t_E,q_0),t_E^+,t_{\mu^+}^+,t_E^-,t_{\mu^-}^+\right)\Theta(\muFi-q_0)  \nonumber\\
&+ \int_0^{E_\chi-m_\chi} q_0^{k+1} dq_0 \,\, \mathcal{I}\left(\tilde{f}_2(t_E,q_0),t_E^+,t_{\mu^-}^-,t_E^-,t_{\mu^+}^-\right)\Theta(\muFi-q_0)  \nonumber\\
&\left. + \int_0^{E_\chi-m_\chi} q_0^{k+1} dq_0 \,\, \mathcal{I}\left(\tilde{f}_2(t_E,q_0),t_E^+,t_{\mu^+}^+,t_E^-,t_{\mu^+}^-\right)\Theta(q_0-\muFi) \right], 
\end{align}
where
\begin{equation}
 t_E^\pm = 2\left[E_\chi(E_\chi-q_0)-m_\chi^2\pm k\sqrt{(E_\chi-q_0)^2-m_\chi^2}\right],    
\end{equation}
\begin{equation}
t_{\mu^{+}}^\pm = 2\left[\muFi(\muFi+q_0)+m_i(2\muFi+q_0)  \pm \sqrt{\left(\muFi(\muFi+q_0)+m_i(2\muFi+q_0)\right)^2-m_i^2q_0^2}\right], 
\end{equation}
\begin{equation}
t_{\mu^{-}}^\pm =  2\left[\muFi(\muFi-q_0)+m_i(2\muFi-q_0)\pm\sqrt{\left(\muFi(\muFi-q_0)+m_i(2\muFi-q_0)\right)^2-m_i^2q_0^2}\right], 
\end{equation}
for further details on the calculation of $t_E^\pm$ and $t_{\mu^\pm}^\pm$ see appendix B of ref.~\cite{Bell:2020jou}.

\section{Interaction rate for up-scattering}
\label{sec:intevaprate}
We perform a similar calculation to that in the previous section, assuming that $q_0$ is now negative.
The derivation of the up-scattering interaction rate is essentially the same as for down scattering until we arrive at the point of the identification of the three different regimes in the response function. For $q_0<0$ these read
\begin{eqnarray}
&E_i&>\muFi-q_0,\\
\mu_{F,n}-q_0>&E_i&>\muFi,\\
&E_i&<\muFi.
\end{eqnarray}
We consider finite values of $\Tstar$, and take the leading contribution, i.e. the terms  of order $e^{-|q_0|/\Tstar}$. For matrix elements independent of $s$ we have
\begin{eqnarray}
\lim_{\Tstar\rightarrow0} \Tstar F_0(E_i/\Tstar,q_0/\Tstar) &=& -q_0 \frac{e^{-|q_0|/\Tstar}}{1-e^{-|q_0|/\Tstar}}, \qquad \qquad \quad E_i>\muFi-q_0,\\
\lim_{\Tstar\rightarrow0} \Tstar F_0(E_i/\Tstar,q_0/\Tstar) &=& (E_i-\muFi)\frac{e^{-|q_0|/\Tstar}}{1-e^{-|q_0|/\Tstar}}, \qquad \muFi-q_0>E_i>\muFi,\\
\lim_{\Tstar\rightarrow0} \Tstar F_0(E_i/\Tstar,q_0/\Tstar) &=& 0, \qquad\qquad\qquad \qquad \qquad \qquad  E_i<\muFi. 
\end{eqnarray}
As in the previous section and appendix~B of ref.~\cite{Bell:2020jou}, we can define
\begin{eqnarray}
\lim_{\Tstar\rightarrow0} \Tstar F_0(E_i/\Tstar,q_0/\Tstar) &=& 
-q_0\frac{e^{q_0/\Tstar}}{1-e^{q_0/\Tstar}} \,  h_0\left(\frac{E_i-\muFi}{q_0}\right),
\end{eqnarray}
and the response function is now given by
\begin{eqnarray}
S^{-}(q_0,q,\Tstar) &=& 
-\frac{e^{q_0/\Tstar}}{1-e^{q_0/\Tstar}} \frac{m_i^2q_0}{\pi q}h_0\left(\frac{E_i^{\,t^{-}}-\muFi}{q_0}\right),\label{eq:SminusEVAP}
\end{eqnarray}
which is equal to Eq.~B.23 of ref.~\cite{Bell:2020jou} times a factor that depends on the temperature and the momentum transfer $q_0$ (and implicitly on $\muFi$ and $B$), 
\begin{equation}
   S^{-}(q_0,q,\Tstar)= -\frac{e^{q_0/\Tstar}}{1-e^{q_0/\Tstar}} S^{-}(q_0,q). 
\end{equation}
Therefore, the differential interaction rate for up-scattering can be estimated with the following expression
\begin{equation}
\dfrac{d\Gamma_{+}^-}{dq_0}\left(q_0,\Tstar\right) = -\frac{e^{q_0/\Tstar}}{1-e^{q_0/\Tstar}} \dfrac{d\Gamma^-}{dq_0}\left(q_0\right),\quad q_0<0.
\end{equation}
 One can then integrate over the variable $t_E$, as in the previous section, to obtain the interaction rate.


\label{Bibliography}

\lhead{\emph{Bibliography}} 

\bibliography{Bibliography} 

\end{document}